# Free-energy calculations for semi-flexible macromolecules: Applications to DNA knotting and looping

Stefan M. Giovan,[1)] Robert G. Scharein,[4)] Andreas Hanke,[5)] and Stephen D. Levene[1,2,3,a)]

*Departments of [1]Molecular and Cell Biology, [2]Bioengineering, and [3]Physics, University of Texas at Dallas, Richardson, TX 75083 USA*

[4]*Hypnagogic Software, Vancouver BC, Canada*

[5]*Department of Physics and Astronomy, University of Texas at Brownsville, Brownsville, TX 78520 USA*

We present a method to obtain numerically accurate values of configurational free energies of semiflexible macromolecular systems, based on the technique of thermodynamic integration combined with normal-mode analysis of a reference system subject to harmonic constraints. Compared with previous free-energy calculations that depend on a reference state, our approach introduces two innovations, namely the use of internal coordinates to constrain the reference states and the ability to freely select these reference states. As a consequence, it is possible to explore systems that undergo substantially larger fluctuations than those considered in previous calculations, including semiflexible biopolymers having arbitrary ratios of contour length $L$ to persistence length $P$. To validate the method, high accuracy is demonstrated for free energies of prime DNA knots with $L/P = 20$ and $L/P = 40$, corresponding to DNA lengths of 3000 and 6000 base pairs, respectively. We then apply the method to study the free-energy landscape for a model of a synaptic nucleoprotein complex containing a pair of looped domains, revealing a bifurcation in the location of optimal synapse (crossover) sites. This transition is relevant to target-site selection by DNA-binding proteins that occupy multiple DNA sites separated by large linear distances along the genome, a problem that arises naturally in gene regulation, DNA recombination, and the action of type-II topoisomerases.

---

[a)] Author to whom correspondence should be addressed. Electronic mail: sdlevene@utdallas.edu.

## I. INTRODUCTION

Free-energy changes govern the direction of all chemical and biophysical processes in biological systems. A quantitative treatment of free-energy landscapes is central to understanding protein and RNA folding,[1] motion and energy transduction in molecular machines,[2] macromolecule-ligand interactions,[3,4] genome organization,[5,6] and many other biological phenomena. For macromolecular systems, obtaining accurate estimates of the free energy is one of the most challenging problems in computational biology and chemistry.[7-9] Systems involving intermediate length scales such as semi-flexible DNA-protein structures are abundant in living cells; however, this class of problems has generally eluded standard computational free-energy methods because the inherent flexibility of molecules on these elevated length scales results in large conformational fluctuations in rugged potential energy landscapes.[10]

The free energy, $F$, of a system in the canonical ensemble, *i.e.*, at constant number of particles $N$, volume $V$, and temperature $T$, is defined as

$$F = \langle U \rangle - TS = -k_B T \ln(Q) \quad , \tag{1}$$

where $\langle U \rangle$ is the mean energy, $S$ is the entropy, and $Q$ is the canonical partition function of the system ($k_B$ is Boltzmann's constant). For simplicity we restrict ourselves to a classical system of $N$ point particles with Cartesian position coordinates $\mathbf{r}_i$, $i = 1,\ldots,N$ in a volume $V$. Assuming that the kinetic energy of the system is independent of the particle positions, $Q$ reduces to the configuration integral

$$Q = \int_V \frac{d^3 r_1}{a^3} \cdots \int_V \frac{d^3 r_N}{a^3} \exp\left[-\beta U(\bar{r})\right] \quad , \tag{2}$$

where $\beta = (k_B T)^{-1}$ and $U(\bar{r})$ is the total potential energy (also referred to as force field) for a full-system configuration $\bar{r} = (\mathbf{r}_1,\ldots,\mathbf{r}_N)$. The constant $a$ in Eq. (2) is a microscopic length required to make $Q$ dimensionless; *e.g.*, for a system of point particles of mass $m$ undergoing Newtonian dynamics, the length $a$ corresponds to the thermal wavelength, $\Lambda = h/(2\pi m k_B T)^{1/2}$, where $h$ is Planck's constant. Equilibrium ensembles of full-system configurations distributed according to the Boltzmann distribution $p(\bar{r}) \propto \exp\left[-\beta U(\bar{r})\right]$ can be generated by using molecular dynamics, Markov-chain Monte Carlo, and Langevin dynamics simulations, for example. To enhance sampling, various schemes have been employed such as parallel tempering (replica-exchange sampling), multicanonical (flat histogram) sampling, and umbrella sampling.[11-15]

A typical application of computational free-energy methods is to compute the *difference* in free energy between two macromolecular states $A$, $B$, i.e.,

$$\Delta F_{AB} = F_B - F_A = -k_B T \ln\left(\frac{Q_B}{Q_A}\right). \qquad (3)$$

If the two states coexist in a simulation, $\Delta F_{AB}$ can be estimated directly by using a counting method, because the ratio of partition functions $Q_B/Q_A$ in Eq. (3) is equal to the probability ratio $P_B/P_A$ of observing states $B$ and $A$, respectively. Thus, counting the respective numbers of times the system visited states $A$ and $B$ during the simulation, $\Omega_A$, $\Omega_B$, and using $Q_B/Q_A = \Omega_B/\Omega_A$ in Eq. (3), yields $\Delta F_{AB}$. In cases where the states $A$ and $B$ do not coexist, alternative sampling methods based on free energy perturbation can be used, such as thermodynamic integration.[7-9] These approaches have been used to compute solvation free energies,[16,17] ligand-receptor binding affinities,[18-20] free energies of protein and RNA folding,[21-24] and in a number of other applications. Thermodynamic integration is used when the two states $A$, $B$ are significantly different from one another, and amounts to sub-dividing (staging) the transformation from $A$ to $B$ into a path of intermediates along a suitable reaction coordinate. Since $\Delta F_{AB}$ is independent of the path between the terminal states $A$, $B$, the intermediates do not have to be physically realistic, which implies that computationally favorable paths may be chosen; however, in the rugged potential energy landscape $U(\vec{r})$ typical of macromolecular systems the task of finding the most favorable paths is difficult in general.[25]

Another class of methods estimates the *absolute* free energy of a macromolecular system for a given force field $U(\vec{r})$. For small systems, limited to harmonic fluctuations about a mechanical-equilibrium ground state, the absolute free energy can be computed by treating fluctuations harmonically about an energy-minimized ground state (the harmonic approximation, or HA), normal mode analysis,[26-28] or use of the so-called quasi-harmonic approximation (QHA).[29,30] However, for larger systems with multiple occupied energy wells, QHA tends to significantly overestimate the configurational entropy.[30,31] To overcome this limitation, the method has been extended by applying QHA to each local minimum (basin) of the potential-energy landscape separately.[32,33] Since the number of basins to be considered increases exponentially with system size, such minima-mining approaches have been limited to protein subdomains and other small systems.[10] Other methods for estimating absolute free energies of macromolecules are based on expressing the full-system probability distribution $p(\vec{r})$ in terms of distributions of molecular fragments, by using an expansion of mutual information terms[34] and polymer growth models.[35-37] In the latter approach, new monomers are added one at a time to an ensemble of partially grown configurations while keeping track of appropriate statistical weights. These techniques have been applied to simple model proteins,[38] RNA secondary structures,[39] all-atom models of peptides and small protein subdomains,[9,40,41] and a number of other systems (see also[42]). A general challenge for polymer growth methods, which becomes more severe for increasing system size, is that configurations important in the full system may have low statistical probability in early stages of growth, so that biasing toward structural information known for the full system is required.[43]

A widely used strategy to estimate the absolute free energy of a given target system, effectively combining methods for relative and absolute free energies described above, employs a reference system, "ref", for which the absolute free energy, $F_{ref}$, is available.[11,44-47] The free energy of the target system, $A$, is calculated as

$$F_A = F_{ref} + \Delta F_{ref \to A} \qquad (4)$$

where $\Delta F_{ref \to A}$ is the free-energy difference between the reference system and the target system, which can be estimated, *e.g.*, by free-energy perturbation or thermodynamic integration. This strategy was introduced for a molecular system using a reference state constrained by external harmonic wells.[44] Using Eq. (4) for two arbitrary target states, $A$, $B$, yields a path-independent method to calculate the free-energy difference $\Delta F_{AB} = F_B - F_A$ in terms of a thermodynamic cycle, without the need of transforming one state to the other.[45] In reference,[46] rather than using harmonic constraints, the reference state was constructed by generating histograms of the coordinates of the target system obtained in a finite-time simulation, and choosing an ensemble of reference configurations at random from the histograms. This approach was later generalized to larger molecules by using pre-calculated molecular fragments as reference states.[47]

In the limit of long, flexible (synthetic) polymer chains powerful tools in statistical physics, such as scaling approaches and the renormalization group, have been widely used to study equilibrium properties of these systems.[48-53] Likewise, the free-energy change associated with *spatially confining* a long, semi-flexible polymer chain with contour length $L$ and persistence length $P$, *e.g.*, into a cylindrical tube or spherical cavity of diameter $D$, have been studied in the asymptotic limit $L \gg P, D$ by combining scaling approaches with Monte Carlo simulations.[54,55] These approaches thus complement methods applicable to the small systems described above. However, semi-flexible macromolecular systems involving intermediate length scales remain notoriously difficult to handle, because such systems are in general out of reach of methods developed for the limiting cases of small, stiff systems and large, flexible systems, respectively.

Critical aspects of DNA replication,[56] recombination,[57,58] repair,[59,60] and gene regulation[61-64] depend on interactions between DNA-bound protein molecules that are separated by large linear distances along the genome, entailing the formation of DNA or chromatin loops. For more than two decades, many details of these processes have been worked out from *in-vitro* and *in-vivo* model studies using plasmid DNAs. The use of circular DNA has been an especially powerful tool in elucidating the mechanisms of recombinases and topoisomerases through topological analysis of knotted and linked products that are trapped by loop formation.[65-68] Loop formation in circular DNA is distinct from that in linear molecules; even in the simplest case of a single pair of interacting sites on circular DNA, there is an excess entropy loss relative to that for forming a pair of independent loops.[69] Thus, the thermodynamics of loop formation within a circular domain has remained an unsolved problem for semi-flexible systems. Combined with recent computational and experimental results suggesting that DNA-loop-mediated processes are driven thermodynamically rather than kinetically,[70-72] there is strong motivation to develop new methods for more generally evaluating the free energies of looped, semiflexible DNA structures.

In this work, we present a method to obtain configurational free energies of semiflexible macromolecular systems by combining thermodynamic integration (TI) with normal mode analysis (NMA). Following the strategy in Eq. (4), an arbitrary molecular state $A$ is gradually transformed into a harmonically constrained reference state $A_0$ and the associated change in free energy is computed by TI. The free energy of the reference state $A_0$ is then computed separately and accurately by means of NMA. Compared with previous free-energy calculations that depend on the use of a reference state,[11,44-46,73] our method introduces two innovations, namely the use of internal coordinates to harmonically constrain the reference states, and the ability to freely select the reference states. As a consequence, it is now possible to explore systems that undergo substantially larger fluctuations than those considered in previous calculations. In addition, using freely selectable reference states avoids problems associated with locating the actual minimum of a rugged potential-energy landscape.[10] The general procedure is numerically accurate, free of uncontrolled approximations, and applies to models of any macromolecular system for which internal geometric constraints can be specified. In particular, our method is applicable to biopolymers having arbitrary ratios of contour length $L$ to persistence length $P$. As a test of our method we show that free energies of prime DNA knots containing up to six crossings can be accurately computed for large, semiflexible DNAs 3000 and 6000 base pairs in size (Fig. 1).

We then apply our method to an open problem, namely investigating the free-energy landscape of circular DNA molecules partitioned into two looped domains by a nucleoprotein complex bound simultaneously to two DNA sites (Fig. 2). These structures are appropriate models for double-strand-passage intermediates in type-II topoisomerase reactions,[74] synaptic complexes in recombination reactions taking place on DNA circles,[68] and promoter selection in enhancer-multiple promoter interactions,[61,62,75-77] among other systems. For a circle of total contour length $L$, synapsis between sites separated by a linear distance $\ell$ generates looped domains with lengths $\ell$ and $L-\ell$, respectively. By evaluating the free-energy landscape $F(\ell, L; P)$ we find that the location of optimal synapse sites on a circular DNA molecule depends strongly on the ratio $L/P$; in particular, we find a bifurcation in the location of optimal synapse sites at a critical value $L/P \approx 10.7$ (about 1600 base pairs for $P = 50$ nm).

## II. METHODS

### A. DNA model: Extensible harmonic chain

We consider a semi-flexible harmonic chain as a coarse-grained mesoscopic model for duplex DNA.[78,79] Chain elements are extensible, cylindrical segments with equilibrium length $b_0$ and fixed diameter $d$, connected end-to-end by semi-flexible joints located at vertices $\mathbf{r}_i$, $i = 1, \ldots, N$ (Fig. 1A). Segments are described by displacement vectors $\mathbf{b}_i = \mathbf{r}_{i+1} - \mathbf{r}_i$ with length $b_i$ and unit-length direction vectors $\hat{\mathbf{b}}_i = \mathbf{b}_i / b_i$. The total potential energy for a given conformation $\bar{r} = (\mathbf{r}_1, \ldots, \mathbf{r}_N)$ is defined as

$$U(\bar{r}) = U_{el} + U_{ev} + U_K, \qquad (5)$$

where $U_{el}$ is the elastic energy of the chain (cf. Eq. (6) below) and $U_{ev}$, $U_K$ describe infinite potential barriers associated with excluded volume (overlapping chain segments) and changes in knot type $K$, respectively (cf. Supplemental Material[80]). These contributions are $U_{ev} = 0$ if none of the cylinders overlap and $U_{ev} = \infty$ otherwise. Similarly, $U_K = 0$ if the chain has knot type $K$ and $U_K = \infty$ otherwise.

In this work we consider two distinct topological models, namely knotted, circular DNA and an unknotted, dual-loop synaptic nucleoprotein complex. Knotted, circular DNA with overall contour length $L$ is modeled as a chain of $N$ segments (Fig. 1). Such knotted chains are formed, *e.g.*, through random closure of linear DNA or via strand passage mediated by topoisomerases or site-specific recombinases. The closure (boundary) condition for circular chains is enforced by the constraint $\mathbf{r}_1 \equiv \mathbf{r}_{N+1}$. In this work we consider the elastic energy due to bending and stretching of the chain only, as appropriate for nicked DNA. The elastic energy for a conformation $\bar{r}$ is defined as

$$U_{el}(\bar{r}) = U_{sc} + k_B T \sum_{i=1}^{N} \left[ c_b \left(1 - \hat{\mathbf{b}}_i \cdot \hat{\mathbf{b}}_{i+1}\right) + \frac{c_s}{2} \left(\frac{b_i}{b_0} - 1\right)^2 \right] , \qquad (6)$$

where $c_b$, $c_s$ are bending and stretching elastic constants, respectively. Elastic energy constants are set so that each segment represents approximately 30 base pairs of DNA (segment length $b_0 = 0.2P$) in 0.15 M NaCl (cf. Supplemental Material[80]). The term $U_{sc}$ in Eq. (6) represents the elastic energy associated with deformations of the synaptic complex (sc) (cf. Eq. (7) below) and does not contribute to the energy of unlooped circular chains.

Dual-loop synaptic complexes consist of a circular DNA molecule having two loci bound to a single protein complex (Fig. 2). The circular contour with length $L$ is therefore partitioned into domains of length $\ell$ and $L - \ell$, respectively, where $\ell$ is the contour length of one of the loops. The loops are modeled as chains having $n$ and $N - n$ segments, respectively, where $N$ is the total number of chain segments and $n$ is the number of segments in the loop of length $\ell$. Both loops are connected at a common vertex $\mathbf{r}_1$, which joins 4 segments $(\mathbf{b}_1, \mathbf{b}_n, \mathbf{b}_{n+1}, \mathbf{b}_N)$ at a four-way junction. The closure (boundary) condition at the junction is enforced by the constraint $\mathbf{r}_1 \equiv \mathbf{r}_{n+1} \equiv \mathbf{r}_{N+1}$. The elastic energy of the synaptic complex (sc) is given by (Fig. 2)

$$U_{sc} = k_B T \frac{c_b}{2} \left( \hat{\mathbf{b}}_1 \cdot \hat{\mathbf{b}}_n + \hat{\mathbf{b}}_1 \cdot \hat{\mathbf{b}}_{n+1} + \hat{\mathbf{b}}_n \cdot \hat{\mathbf{b}}_N + \hat{\mathbf{b}}_{n+1} \cdot \hat{\mathbf{b}}_N \right) . \qquad (7)$$

Thus, the preferred geometry of the synapse in this model is that of a square-planar crossing, which has been proposed as a prototype Holliday-junction geometry in the phage-λ integrase superfamily of site-specific recombinases.[81-83]

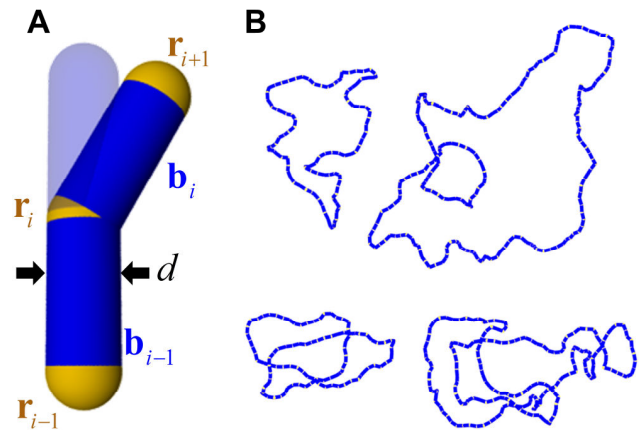

FIG. 1. Coarse-grained DNA model. (A) Definition of chain segments $\mathbf{b}_i$, vertices $\mathbf{r}_i$, and segment diameter $d$. The equilibrium position of segment $\mathbf{b}_i$ is shown by the translucent surface. (B) Typical chain conformations generated by Monte Carlo simulation. Shown are chains with $N = 100$ segments ($L = 20P$, 3000 base pairs) (left) and chains with $N = 200$ segments ($L = 40P$, 6000 base pairs) (right). The top chains are unknotted and the bottom chains are knotted ($4_1$, figure-eight).

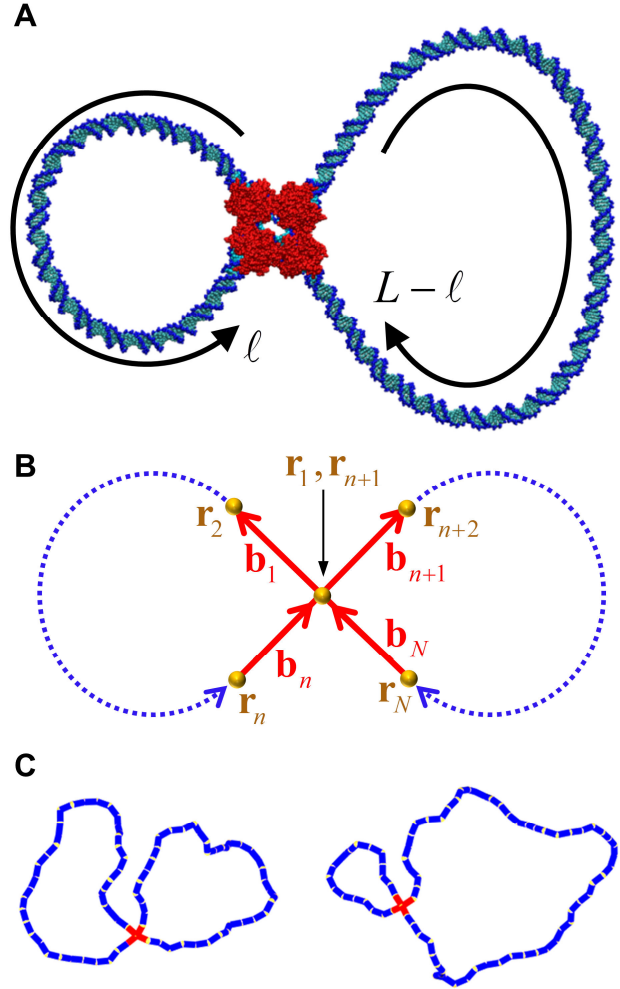

FIG. 2. Dual-loop synaptic complex. (A) Atomistic model of a nucleoprotein synaptic complex (Cre-LoxP, adapted from PDB ID 3CRX) formed on a circular DNA substrate. Cre-protein atoms are shown in red. (B) Definition of segment vectors forming the synaptic junction (see text for details). (C) Typical chain conformations generated by Monte Carlo simulation. Both chains have the same total length $L = 16P$ (2400 base pairs), but different locations of the synapse (red chain segments). At left, the synapse is located at $\ell = L/2$, forming two loops of equal length $\ell = 8P$. At right, the synapse is located at $\ell = L/4$, forming loops with lengths $\ell = 4P$ and $L - \ell = 12P$, respectively.

**B. Free-energy calculations using TI-NMA**

We used a Metropolis Monte Carlo procedure to generate equilibrium ensembles of chain conformations (cf. Supplemental Material[80]). Free energies of knotted, circular chains and dual-loop synaptic complex models were obtained by combining Thermodynamic Integration (TI) with Normal Mode Analysis (NMA), a scheme henceforth referred to as TI-NMA. NMA requires the system to obey harmonic fluctuations about a well-defined minimum-energy configuration. To this end, we use an auxiliary potential energy function, $U_{ha}(\vec{r},\vec{r}_0)$, which exhibits a single, well-defined minimum at $\vec{r}_0$ such that $U_{ha}(\vec{r}=\vec{r}_0) \equiv 0$. The ground-state conformation $\vec{r}_0$ may be selected arbitrarily (cf. Section II.C. below). To calculate $U_{ha}(\vec{r},\vec{r}_0)$ we associate basis vectors $\{\hat{\mathbf{x}}_i,\hat{\mathbf{y}}_i,\hat{\mathbf{b}}_i\}$ with all segments $\mathbf{b}_i$ as follows. Position vectors $\mathbf{r}_i$ are given relative to the chain's center of mass, $\mathbf{R}_{cm} = N^{-1}\sum_i^N \mathbf{r}_i$, where each vertex of the chain at position $\mathbf{r}_i$ is taken to have unit mass. The $z$-axes of the frames are defined as the unit-length direction vectors $\hat{\mathbf{b}}_i$, $y$-axes are defined as $\hat{\mathbf{y}}_i = \mathbf{r}_i \times \hat{\mathbf{b}}_i / |\mathbf{r}_i \times \hat{\mathbf{b}}_i|$, and $x$-axes are defined as $\hat{\mathbf{x}}_i = \hat{\mathbf{y}}_i \times \hat{\mathbf{b}}_i$ (Fig. 3). The potential $U_{ha}$ is then defined as

$$U_{ha}(\vec{r},\vec{r}_0) = U_{sc} + k_B T \sum_{i=1}^{N}\left[\frac{c_b}{2}\left(2 - \hat{\mathbf{b}}_{i+1}\cdot\hat{\mathbf{b}}^p_{i+1} - \hat{\mathbf{b}}_{i-1}\cdot\hat{\mathbf{b}}^p_{i-1}\right) + \frac{c_s}{2}\left(\frac{b_i}{b_0}-1\right)^2\right], \qquad (8)$$

where $\hat{\mathbf{b}}^p_{i+1}$, $\hat{\mathbf{b}}^p_{i-1}$ are the preferred orientations of segment vectors adjacent to $\mathbf{b}_i$ in frame $\{\hat{\mathbf{x}}_i,\hat{\mathbf{y}}_i,\hat{\mathbf{b}}_i\}$. The preferred segment directions are given by the orientations of segments $\hat{\mathbf{b}}^0_{i+1}$, $\hat{\mathbf{b}}^0_{i-1}$ within frames $\{\hat{\mathbf{x}}^0_i,\hat{\mathbf{y}}^0_i,\hat{\mathbf{z}}^0_i\}$ of the reference configuration $\vec{r}_0$ (cf. Section II.C. below).

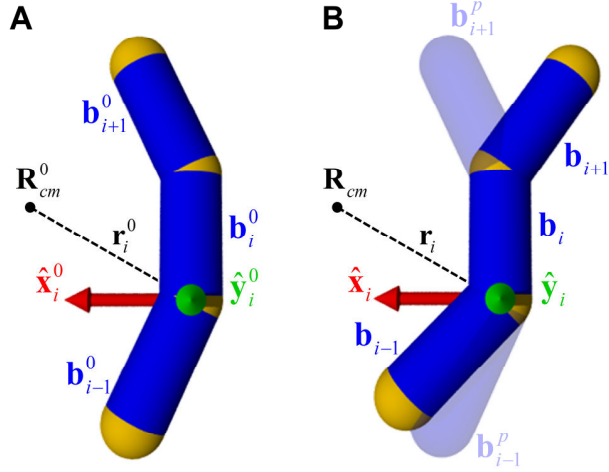

FIG. 3. Internal-coordinate constraints. (A) For a given ground state conformation $\vec{r}_0$ with center of mass $\mathbf{R}_{cm}^0$, the basis frames $\{\hat{\mathbf{x}}_i^0, \hat{\mathbf{y}}_i^0, \hat{\mathbf{b}}_i^0\}$ are obtained for each segment $\mathbf{b}_i^0$ such that axis $\hat{\mathbf{x}}_i^0$ lies in the plane defined by the point $\mathbf{R}_{cm}^0$ and the vector $\mathbf{b}_i^0$, and points towards $\mathbf{R}_{cm}^0$. (B) For an arbitrary conformation $\vec{r}$, frames $\{\hat{\mathbf{x}}_i, \hat{\mathbf{y}}_i, \hat{\mathbf{b}}_i\}$ are obtained in the same way as in (A). The equilibrium (preferred) orientations $\mathbf{b}_{i+1}^p$, $\mathbf{b}_{i-1}^p$ (shown as translucent surfaces) of segments $\mathbf{b}_{i+1}$, $\mathbf{b}_{i-1}$ in the frame $\{\hat{\mathbf{x}}_i, \hat{\mathbf{y}}_i, \hat{\mathbf{b}}_i\}$ correspond to the orientations $\mathbf{b}_{i+1}^0$, $\mathbf{b}_{i-1}^0$ in the frame $\{\hat{\mathbf{x}}_i^0, \hat{\mathbf{y}}_i^0, \hat{\mathbf{b}}_i^0\}$.

The objective is to replace the energy function of the original, semi-flexible system described by Eq. (5) with a potential function suitable for NMA, and to calculate the associated change in free energy by TI (Fig. 4). To this end we define

$$U(\lambda) = \begin{cases} \lambda U_{ha} + (1-\lambda) U_{el} + U_{ev} + U_K, & 0 \leq \lambda \leq 1 \\ \lambda U_{ha} + U_{ev} + U_K, & 1 \leq \lambda \leq \lambda_{max} \end{cases} \quad (9)$$

where the parameter $\lambda$ serves to switch between the limits $U_{el} + U_{ev} + U_K$ at $\lambda = 0$ and $\lambda_{max} U_{ha} + U_{ev} + U_K$ at $\lambda = \lambda_{max}$. The first phase of TI, $0 \leq \lambda \leq 1$, replaces the original elastic potential energy $U_{el}(\vec{r})$ with the new potential $U_{ha}(\vec{r}, \vec{r}_0)$. The second phase of TI further increases $\lambda$ to a value $\lambda_{max}$ which is chosen large enough so that the system is fully constrained to harmonic fluctuations about the reference ground state $\vec{r}_0$. This process is shown for selected knotted and dual-loop sc models in Supplemental Material[80] (Movies 1a-j). Finally, NMA is applied to the ground state configuration $\vec{r}_0$ and the total free energy $F$ is obtained as

$$F = F_0 - \Delta F_1 - \Delta F_2 \ . \tag{10}$$

$\Delta F_1$ and $\Delta F_2$ are the free-energy changes associated with the two phases of TI described above and $F_0$ is the free energy of the reference ground state $\vec{r}_0$ obtained by NMA. Selected normal modes for knotted and dual-loop sc models are shown in Supplemental Material[80] (Movies 2a-j). Carrying out this procedure for two different semi-flexible systems $A$, $B$, e.g., two different DNA knots, yields the free-energy difference $\Delta F_{AB} = F_B - F_A$ in terms of the thermodynamic cycle shown in Fig. 4.

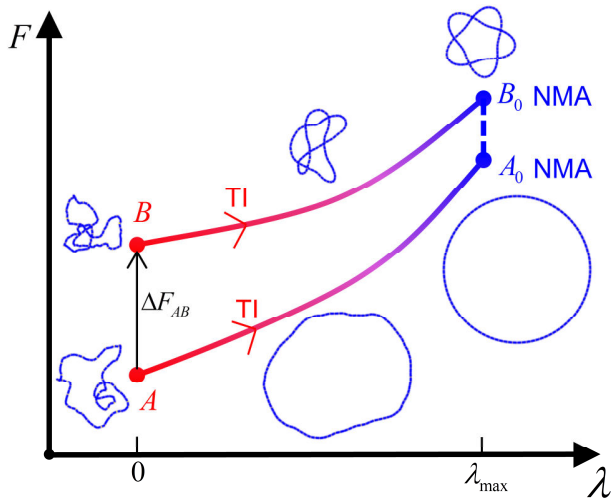

FIG. 4. Thermodynamic cycle yielding the free-energy difference between states $A$ and $B$ (red section of the curves). Thermodynamic integration (TI) computes the change in free energy as each of the states $A$ (unknotted), $B$ (knotted, $5_1$) is transformed into ground-state conformations $A_0$, $B_0$ (blue section of the curves). Normal mode analysis (NMA) gives the free energy of the ground-state conformations for the states $A_0$, $B_0$, completing the cycle (dashed blue line). Typical Monte Carlo conformations are shown for the TI portion of the cycle. Ground-state conformations of unknotted and knotted chains used for NMA are shown on the far right.

## C. Reference systems

A major advantage of our method is the ability to freely select, in principle, any arbitrary full-system configuration as a possible reference state. These states can be chosen on the basis of computational convenience, the final computed difference in free energy being independent of the reference states. For the present calculations, reference conformations $\vec{r}_0$ for knotted chains were idealized knot conformations having unit-length segments, which were generated by the program KnotPlot.[84] Conformations $\vec{r}_0$ for the dual-loop synapse model were obtained by gradient-descent energy minimization of $U_{el}$ in Eq. (6). The reference conformations used in our calculations are shown in Supplemental Material[80] (Fig. S2). In what follows we assign preferred segment orientations $\hat{\mathbf{b}}_{i+1}^p$, $\hat{\mathbf{b}}_{i-1}^p$ for the segment vectors adjacent to $\mathbf{b}_i$ to be used in the auxiliary potential energy function $U_{ha}(\vec{r},\vec{r}_0)$ in Eq. (8). At first, segment frames $\{\hat{\mathbf{x}}_i^0,\hat{\mathbf{y}}_i^0,\hat{\mathbf{b}}_i^0\}$ are obtained for the chosen ground state configuration $\vec{r}_0$ (cf. Fig. 3A). Displacement vectors $\hat{\mathbf{b}}_{i+1}^0$, $\hat{\mathbf{b}}_{i-1}^0$ are expanded in the frame $\{\hat{\mathbf{x}}_i^0,\hat{\mathbf{y}}_i^0,\hat{\mathbf{b}}_i^0\}$ with expansion coefficients

$$b_{i+1,x}^0 = \hat{\mathbf{b}}_{i+1}^0 \cdot \hat{\mathbf{x}}_i^0 \ , \ b_{i+1,y}^0 = \hat{\mathbf{b}}_{i+1}^0 \cdot \hat{\mathbf{y}}_i^0 \ , \ b_{i+1,z}^0 = \hat{\mathbf{b}}_{i+1}^0 \cdot \hat{\mathbf{b}}_i^0 \ , \tag{11}$$

$$b_{i-1,x}^0 = \hat{\mathbf{b}}_{i-1}^0 \cdot \hat{\mathbf{x}}_i^0 \ , \ b_{i-1,y}^0 = \hat{\mathbf{b}}_{i-1}^0 \cdot \hat{\mathbf{y}}_i^0 \ , \ b_{i-1,z}^0 = \hat{\mathbf{b}}_{i-1}^0 \cdot \hat{\mathbf{b}}_i^0 \ . \tag{12}$$

The above $6N$ parameters are then used to define preferred orientations $\hat{\mathbf{b}}_{i+1}^p$, $\hat{\mathbf{b}}_{i-1}^p$ for an arbitrary configuration $\vec{r}$, with segment frames $\{\hat{\mathbf{x}}_i,\hat{\mathbf{y}}_i,\hat{\mathbf{b}}_i\}$ (cf. Fig. 3B), as

$$\hat{\mathbf{b}}_{i+1}^p = b_{i+1,x}^0 \hat{\mathbf{x}}_i + b_{i+1,y}^0 \hat{\mathbf{y}}_i + b_{i+1,z}^0 \hat{\mathbf{b}}_i \ , \tag{13}$$

$$\hat{\mathbf{b}}_{i-1}^p = b_{i-1,x}^0 \hat{\mathbf{x}}_i + b_{i-1,y}^0 \hat{\mathbf{y}}_i + b_{i-1,z}^0 \hat{\mathbf{b}}_i \ . \tag{14}$$

### D. Thermodynamic Integration (TI)

In the first phase of TI, for $0 \leq \lambda \leq 1$, we replace the elastic potential energy $U_{el}$ by the potential $U_{ha}$ which constrains the chain to a reference ground state $\vec{r}_0$. The free-energy change associated with this process is given by

$$\Delta F_1 = \int_0^1 d\lambda \left\langle \frac{dU}{d\lambda} \right\rangle_\lambda = \int_0^1 d\lambda \left\langle U_{ha} - U_{el} \right\rangle_\lambda \ , \tag{15}$$

where the subscript indicates that evaluation takes place at a specific value of $\lambda$. Values of $\left\langle U_{ha} - U_{el} \right\rangle_\lambda$ for 21 equally spaced values of $\lambda = \{0, 0.05, 0.1, \ldots, 1\}$ were obtained by Monte Carlo simulation and the results were interpolated and integrated according to Eq. (15) (cf. Supplemental Material[80]). In the second phase of TI, for $1 \leq \lambda \leq \lambda_{max}$, we further increased $\lambda$ until the system was fully constrained to harmonic fluctuations about the reference ground state $\vec{r}_0$. We found that $\lambda_{max} = 200$ was sufficient to achieve high numerical accuracy in all calculations presented here; in particular, we verified that $\langle U(\lambda) \rangle_\lambda$, with $U(\lambda)$ given in Eq. (9), converged to $k_B T (3N-6)/2$ as expected for harmonic behavior according to the equipartition theorem.[85] The free-energy change associated with this process is given by

$$\Delta F_2 = \int_1^{\lambda_{max}} d\lambda \left\langle \frac{dU}{d\lambda} \right\rangle_\lambda = \int_1^{\lambda_{max}} d\lambda \langle U_{ha} \rangle_\lambda . \tag{16}$$

Values of $\langle U_{ha} \rangle_\lambda$ for 100 exponentially increasing values of $\lambda$ from 1 to $\lambda_{max}$ were obtained by Monte Carlo simulation and the results were interpolated and integrated according to Eq. (16) (cf. Supplemental Material[80]).

**E. Normal Mode Analysis (NMA)**

The Hessian matrix of mixed second derivatives, $H_{ij} = \beta b_0^2 \frac{\partial^2 U(\vec{r})}{\partial r_i \partial r_j}$, $i, j = 1, \ldots, 3N$, is calculated and diagonalized to obtain eigenvalues $\nu_m$, $m = 1, \ldots, 3N$, representing force constants for each normal mode $m$ (cf. Supplemental Material[80]). When ordered smallest to largest, one finds $\nu_m = 0$ for $m = 1, \ldots, 6$ and $\nu_m > 0$ for $m = 7, \ldots, 3N$. The 6 zero eigenvalues $m = 1, \ldots, 6$ are associated with rigid translations and rotations of the whole system and thus do not incur any energetic cost. The $3N - 6$ nonzero eigenvalues $m = 7, \ldots, 3N$ are associated with internal vibrations of the chain about the minimum energy configuration $\vec{r}_0$. The free energy is obtained as

$$-\frac{F_0}{k_B T} = \rho + \ln(N^{3/2}) + \ln\left(8\pi^2 \sqrt{I_x I_y I_z}\right) + \frac{1}{2} \sum_{m=7}^{3N} \ln\left(\frac{2\pi}{\nu_m}\right), \tag{17}$$

where $I_x$, $I_y$, $I_z$ are the principal moments of inertia of the minimum energy configuration $\vec{r}_0$ in units of $b_0$ (cf. Eq. (8)). The quantity $\rho$ depends on the discretization of the system, but is constant for molecules of the same size (number of segments) and thus does not appear in differences (cf. Supplemental Material[80]). Note that the potential energy of the minimum configuration $\vec{r}_0$ is zero according to the definition of $U_{ha}(\vec{r}, \vec{r}_0)$ in Eq. (8) and is therefore absent in Eq. (17).

## III. RESULTS

### A. Equilibrium distributions of knotted, circular molecules

To verify the numerical accuracy of the TI-NMA method we calculated free energies of DNA prime knots having up to 6 irreducible crossings. Computations were done for two different DNA lengths, namely $L/P = 20$ ($N = 100$ segments) and $L/P = 40$ ($N = 200$ segments), corresponding to nicked-circular DNA molecules 3000 and 6000 base pairs in size (Supplementary Material, Table S1). Our results for the free energies of DNA knots obtained by TI-NMA are compared directly to the knots' probabilities of occurrence in ensembles of chains generated by random segment passage during successive deformations of the chain (Equilibrium Segment Passage, ESP) (Fig. 5). Such ensembles sample equilibrium distributions of knot types.[84,86] We here extend the ESP method to ensembles in which knots with the highest probability of occurrence were excluded during the simulation. Restricted ensembles produce more efficient sampling and thus higher numerical accuracy for knots with low probability, *i.e.*, high free energy (cf. Supplemental Material[80] for details). Figure 5 shows that values of the free energy for DNAs up to 6000 base pairs (40 $P$) obtained by TI-NMA are numerically accurate to within fractions of $k_BT$.

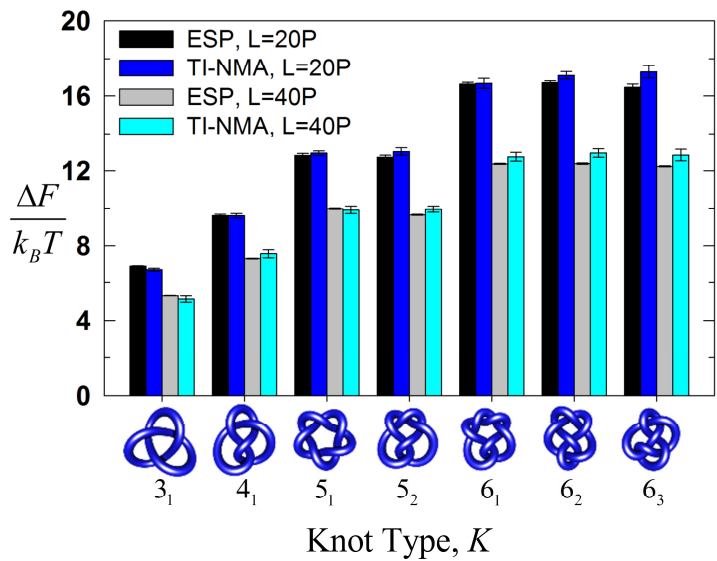

FIG. 5. Knotting free energies of DNA circles relative to an unknot of identical size calculated for different knot types using TI-NMA (blue, cyan bars) and ESP (black, gray bars). Black and blue bars show results for chains with length $L = 20P$ ($N = 100$, 3000 base pairs), whereas gray and cyan bars show results for chains with length $L = 40P$ ($N = 200$, 6000 base pairs).

### 1. Convergence of free-energy values for increasing ensemble size

Absolute free energies of chains with the same overall length but with different knot types were calculated for increasing ensemble sizes (Fig. 6). Simulations were performed at 120 different values of $\lambda$, each producing an ensemble containing $\eta$ independent conformations. The absolute free energy $F = F_0 - \Delta F_{TI}$ was then plotted for increasing $\eta$.

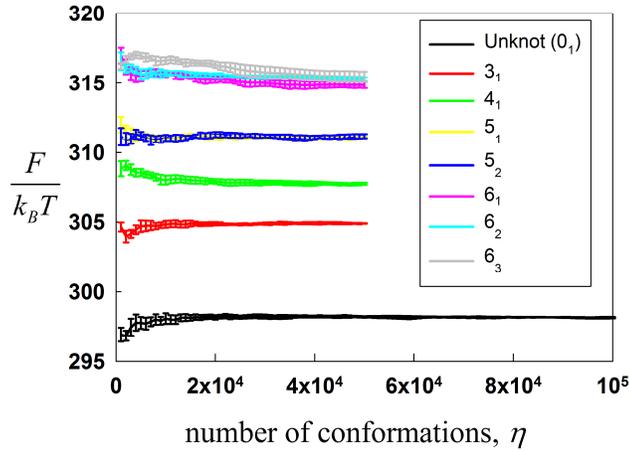

FIG. 6. Free energies of chains with $N = 100$ segments ($L/P = 20$, 3000 base-pairs) are shown for each knot type and for increasing ensemble sizes.

### 2. Independence of the free energy on the reference system

Final computed values for the difference in free energy between two states are independent of the chosen reference states (see Section II.C.). This is demonstrated here by calculating the free energy of a knotted DNA molecule for several arbitrary choices of the reference conformation (Fig. 7). The computed free-energy values are found to coincide within fractions of $k_B T$ (Table 1).

TABLE I. Comparison of free energies obtained for different reference conformations.

| Conformation[a] | $\beta F(\lambda=0)$ [b] | $\beta F_0(\lambda=200)$ | $\beta\Delta F_{TI}(\lambda=0\to 200)$ |
|---|---|---|---|
| a | 308.36(0.16)[c] | 1077.85 | 769.49(0.16) |
| b | 307.94(0.08) | 1076.71 | 768.77(0.08) |
| c | 308.54(0.07) | 1076.45 | 767.91(0.07) |
| d | 308.10(0.14) | 1078.62 | 770.52(0.14) |

[a]cf. Fig. 7

[b] $\beta=(k_B T)^{-1}$

[c]values in parentheses represent error (SEM, 5 trials, $\eta=5\times 10^5$ in each trial)

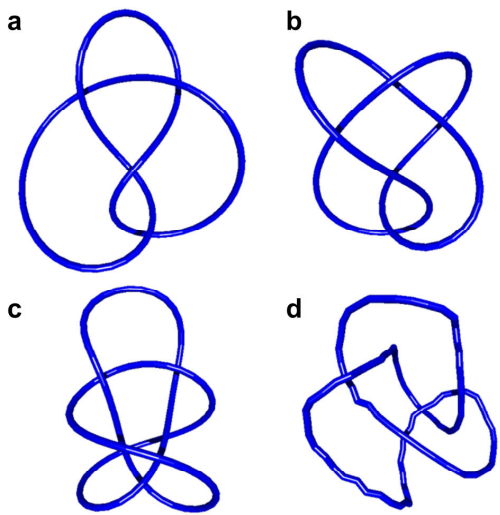

FIG. 7. Reference conformations used to calculate the free energy of a figure-eight (4.1) DNA knot, demonstrating the independence of the computed free energy on chosen reference conformations. Each conformation (a-d) has overall length $L=20P$ ($N=100$ segments). Conformation (a) was used as the reference state for the 4.1 knot with $L=20P$ in Fig. 5. The final free-energy values obtained using these different reference conformations coincide within fractions of $k_B T$ (cf. Table I).

## B. Free-energy landscapes for dual-looped synapse models

We used our approach to compute the free energy landscape $F(\ell,L;P)$ of a synaptic DNA complex containing an internal synapse at two specified DNA segments. $\Delta F_{TI}=\Delta F_1+\Delta F_2$ and $F_0$ were individually fitted to continuous functions of loop length $\ell$ for specific ratios of $L/P$ (cf. Supplemental Material[80] for details). The free energy landscape $F(\ell,L;P)$ was obtained by interpolating these results for intermediate values $L/P$. Figure 8A shows $\Delta F=F(\ell,L;P)-F(L/2,L;P)$ as a function of $\ell/L$ and $L/P$.

The behavior of $\Delta F$ as a function of $\ell$ for fixed $L/P$ depends strongly on the value of $L/P$. For small $L/P$, we find a single minimum for $\Delta F$ at $m = L/2$. At the critical value $L/P \approx 10.7$ the single minimum at $m = L/2$ bifurcates into two separate minima at $m_1 < L/2$ and $m_2 = (L - m_1) > L/2$, respectively. The two minima tend to $m_1/L \to 0$ and $m_2/L \to 1$ as $L/P$ is further increased. The onset of the bifurcation depends on the ratio $L/P$ and may thus be controlled, for fixed length $L$, by varying the persistence length $P$ of the DNA. Figure 8B shows $\Delta F$ along a circular DNA molecule with fixed length $L$ as a function of the separation $\ell$ between synapse sites, where one synapse site is located at the top of each circle. Clearly, the locations of the minima in $\Delta F$, corresponding to optimal locations of synapse sites, changes dramatically as the size of the molecule increases from 1200 to 2100 base pairs.

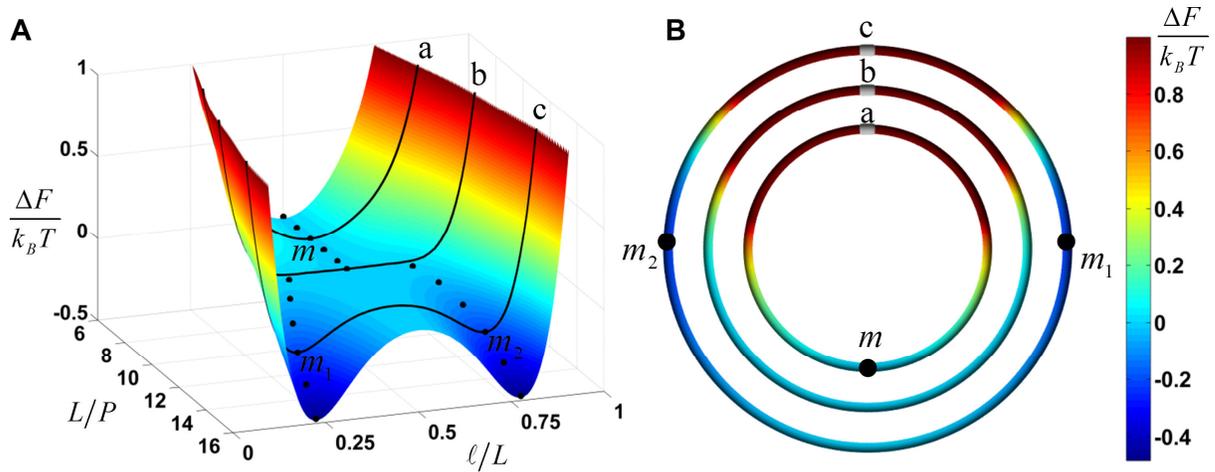

FIG. 8. Free-energy landscape $\Delta F(\ell, L; P)$ of a dual-loop DNA synaptic complex. (A) $\Delta F$ as a function of contour to persistence length ratio, $L/P$, and fractional contour length $\ell/L$ (cf. Fig. 2). At a critical value $L/P \approx 10.7$, the minimum for $\Delta F$ (black dots) bifurcates from a single minimum at $m = L/2$ into two separate minima at $m_1 < L/2$ and $m_2 = (L - m_1) > L/2$, respectively. Black curves labeled **a**, **b**, **c** represent free energy profiles for fixed ratio $L/P$. (B) $\Delta F$ along a circular DNA molecule with fixed ratio $L/P$, corresponding to curves a, b, c in (A), as a function of the separation $\ell$ (to scale). One of the synapse sites is located at the top of each circle. Values of $\Delta F$ are represented by colors as shown. Chain a: $L/P = 8$ (about 1200 base pairs), with a single minimum $m$; chain b: critical value $L/P \approx 10.7$ (about 1600 base pairs); chain c: $L/P = 14$ (about 2100 base pairs), with two separate minima $m_1$ and $m_2$.

## IV. DISCUSSION

We have developed a general approach for estimating the configurational free energy of semi-flexible macromolecules that combines thermodynamic integration (TI) with normal-mode analysis (NMA) of a harmonically constrained reference system. Our approach introduces two important innovations, namely the use of internal coordinates to harmonically constrain reference states, along with the ability to freely select these reference states. As a result, it is now possible to study systems that undergo substantially larger fluctuations than those considered in previous calculations.

By using Eq. (10), the TI-NMA method yields the *absolute* free energy of a macromolecular system for given force field $U(\bar{r})$. Applying Eq. (10) to two arbitrary configurations yields the free-energy difference by means of a thermodynamic cycle, without the need to transform one state into the other (cf. Fig. 4). The TI-NMA method can thereby compute free-energy differences for macromolecular states that are significantly different from one another. We consider one such case here, namely circular DNA molecules of different knot types (cf. Fig. 5). Topological barriers render energy-perturbation methods (including TI) ineffective because there is no well-defined reaction coordinate that continuously connects different topological states. Similar difficulties arise in connection with problems involving phase transitions, such as unfolded and folded states of a protein separated by a first-order collapse transition.[87] Our method is thus applicable to a wide range of problems involving biomolecular organization, beyond free energies of DNA knotting and looping in circular DNA (cf. Figs. 5, 8). The polymer-length scale we address corresponds to that of plasmid-sized DNA circles, a size regime that is highly challenging to other approaches.

In this work, the difference in free energy, $\Delta F$, between reference and target states was computed directly by applying TI to results for $\langle U \rangle_\lambda$ obtained by MC simulation (cf. Eqs. (15), (16)). Figure 6 shows that convergence of the free energy can be achieved with ensembles as small as 20,000 – 40,000 chains; these ensembles are on the order of 100 times smaller than those required by other methods (data not shown). However, in more complex systems, accuracy in $\Delta F$ may require enhanced sampling techniques such as parallel tempering, multicanonical sampling, or umbrella sampling.[11-15]

Applications of the TI-NMA method are particularly appropriate in cases where the topology and/or geometry of a semi-flexible polymer domain is fixed. Examples include topologically closed domains that may be simultaneously knotted and supercoiled. Previous studies assessing the effect of supercoiling on the free energy of knot formation were based on models similar to ours, but led to conflicting conclusions because of differences in imputed changes in DNA supercoiling.[86,88] Such ambiguities can be avoided in our methodology, which can rigorously and systematically address a broad range of problems involving the formation of topologically closed DNA domains.

Our results for the free-energy landscape $F(\ell, L; P)$ of a dual-looped synaptic DNA complex reveal a bifurcation in the optimal location of synapse sites, coinciding with the minima in the free-energy landscape $F(\ell, L; P)$, as the parameter $L/P$ is changed (Fig. 8). The notion of bifurcations was originally introduced in the mathematical study of dynamical systems. In this context, a bifurcation corresponds to a qualitative change in the behavior of a dynamical system when a parameter changes, *e.g.*, the appearance or disappearance of equilibrium points.[89] Similarly, the bifurcation in the location of optimal synapse sites in a synaptic DNA complex found here corresponds to a change in the shape of the free-energy landscape $F(\ell, L; P)$ as a function of $\ell/L$ when the parameter $L/P$ is changed (for example, by changing $P$ for fixed $L$). By identifying and characterizing the bifurcation we are able to address the problem of target-site selection by DNA-binding proteins and protein complexes that occupy multiple DNA sites separated by large linear distances. This problem arises naturally in gene-regulatory contexts that involve interactions between enhancer and multiple promoter sequences, in the action of type-II topoisomerases, and in other sequence-specific and non-specific protein-DNA interactions. We note that topological transitions in Gaussian chains (*i.e.*, without bending energy) constrained by *slip-links*, which are driven purely by entropy upon changing the through-space distance between chain ends, have been known for some time.[90,91] In contrast, the bifurcation in the locations of optimal synapse sites found here results from competition between energetic and entropic contributions to the full free energy of a semi-flexible system.

We note that probing the complete free energy landscape $F(\ell, L; P)$ (Fig. 8) directly by simulating an equilibrium set of conformations of circular chains and counting the number of conformations which happen to have a crosslink at different positions along the chain during the simulation is computationally inefficient. This problem is particularly severe for combinations of parameters $(\ell, L; P)$ for which the free energy $F(\ell, L; P)$ is large (*i.e.*, the probability of occurrence of these states is small). Moreover, the detailed geometric constraints imposed by the synaptic complex on the DNA loops cannot be realistically implemented in such a model. The TI-NMA method avoids this problem by computing the free energy of the whole protein-DNA synaptic complex for individual combinations of parameters $(\ell, L; P)$, taking into account realistic geometric parameters for the complex.

In the asymptotic limit of long, flexible chains, our results for $F(\ell, L; P)$ (Fig. 8) are consistent with the phenomenon of *knot localization* found previously, *i.e.*, the statistically most-probable conformations minimizing $F(\ell, L; P)$ are those for which one of the loop sizes is much smaller than the other.[69,92-95] The TI-NMA method allows us to go beyond this result and to obtain realistic values for the free energy $F(\ell, L; P)$ of semi-flexible polymers with arbitrary ratios $L/P$. Our method can be readily extended to finding optimal synapse spacing as a function of the ionic environment, supercoiling, and locations of additional DNA-binding proteins.

**ACKNOWLEDGMENTS**

We are grateful to Sarah Harris for important discussions. This work was supported by the NSF/NIH Joint Program in Mathematical Biology (NSF grant DMS-0800929) to SDL and NIH grant 2SC3GM083779-04A1 to AH.

**REFERENCES**

[1] S. J. Chen, Annu. Rev. Biophys. **37**, 197 (2008).
[2] A. Y. Lyubimov, M. Strycharska and J. M. Berger, Curr. Opin. Struct. Biol. **21**, 240 (2011).
[3] M. K. Gilson and H. X. Zhou, Annu. Rev. Biophys. Biomol. Struct. **36**, 21 (2007).
[4] F. Bai, Y. Xu, J. Chen, Q. Liu, J. Gu, X. Wang, J. Ma, H. Li, J. N. Onuchic and H. Jiang, Proc. Natl. Acad. Sci. U.S.A. **110**, 4273 (2013).
[5] S. Jun and A. Wright, Nature Rev. Microbiol. **8**, 600 (2010).
[6] C. A. Brackley, S. Taylor, A. Papantonis, P. R. Cook and D. Marenduzzo, Proc. Natl. Acad. Sci. U.S.A. **110**, E3605 (2013).
[7] C. D. Christ, A. E. Mark and W. F. van Gunsteren, Journal of computational chemistry **31**, 1569 (2010).
[8] C. Chipot and A. Pohorille (Eds.), *Free Energy Calculations: Theory and Applications in Chemistry and Biology*. (Springer, Berlin, 2007).
[9] H. Meirovitch, S. Cheluvaraja and R. P. White, Current protein & peptide science **10**, 229 (2009).
[10] D. Wales, *Energy Landscapes: Applications to Clusters, Biomolecules and Glasses*. (Cambridge University Press, Cambridge, 2004).
[11] D. Frenkel and B. Smit, *Understanding Molecular Simulation: From Algorithms to Applications*, 2nd ed. (Academic Press, San Diego, 2002).
[12] B. Berg, in *Rugged Free Energy Landscapes*, edited by W. Janke (Springer, Berlin, 2008), Vol. 736, pp. 317.
[13] D. P. Landau and K. Binder, *A Guide to Monte Carlo Simulations in Statistical Physics*, 3rd ed. (Cambridge University Press, Cambridge, 2009).
[14] D. M. Zuckerman, Annu. Rev. Biophys. **40**, 41 (2011).
[15] R. O. Dror, R. M. Dirks, J. P. Grossman, H. Xu and D. E. Shaw, Annu. Rev. Biophys. **41**, 429 (2012).
[16] M. R. Shirts and V. S. Pande, J. Chem. Phys. **122**, 134508 (2005).
[17] S. Lee, K.-H. Cho, Y.-M. Kang, H. A. Scheraga and K. T. No, Proc. Natl. Acad. Sci. U.S.A. **110**, E662 (2013).
[18] A. Aleksandrov, D. Thompson and T. Simonson, J. Mol. Recognit. **23**, 117 (2010).
[19] H.-J. Woo and B. Roux, Proc. Natl. Acad. Sci. U.S.A. **102**, 6825 (2005).
[20] H. Kokubo, T. Tanaka and Y. Okamoto, J. Chem. Theory Comput. **9**, 4660 (2013).
[21] J. Wang, Q. Shao, Z. Xu, Y. Liu, Z. Yang, B. P. Cossins, H. Jiang, K. Chen, J. Shi and W. Zhu, J. Phys. Chem. B **118**, 134 (2014).
[22] S. Uyaver and U. H. E. Hansmann, J. Chem. Phys. **140**, 065101 (2014).
[23] S.-J. Chen and K. A. Dill, Proc. Natl. Acad. Sci. U.S.A. **97**, 646 (2000).
[24] S. S. Cho, D. L. Pincus and D. Thirumalai, Proc. Natl. Acad. Sci. U.S.A. **106**, 17349 (2009).
[25] D. Branduardi, F. L. Gervasio and M. Parrinello, J. Chem. Phys. **126**, 054103 (2007).
[26] Q. Cui and I. Bahar (Eds.), *Normal mode analysis: theory and applications to biological and chemical systems*. (Chapman & Hall/CRC, Boca Raton, 2006).
[27] Y. Zhang, A. E. McEwen, D. M. Crothers and S. D. Levene, Biophys. J. **90**, 1903 (2006).
[28] N. Clauvelin, W. K. Olson and I. Tobias, J Elast **115**, 193 (2014).
[29] S. A. Harris and C. A. Laughton, J. Phys.: Condens. Matter **19**, 076103 (2007).
[30] U. Hensen, O. F. Lange and H. Grubmuller, PLoS ONE **5**, e9179 (2010).
[31] C. E. Chang, W. Chen and M. K. Gilson, J. Chem. Theory Comput. **1**, 1017 (2005).
[32] T. V. Bogdan, D. J. Wales and F. Calvo, J. Chem. Phys. **124**, 044102 (2006).
[33] C. E. Chang, W. Chen and M. K. Gilson, Proc. Natl. Acad. Sci. U.S.A. **104**, 1534 (2007).
[34] B. J. Killian, J. Yundenfreund Kravitz and M. K. Gilson, J. Chem. Phys. **127**, 024107 (2007).
[35] F. T. Wall and J. J. Erpenbeck, J. Chem. Phys. **30**, 634 (1959).
[36] H. Meirovitch, J. Phys. A: Math. Gen. **15**, L735 (1982).
[37] T. Garel and H. Orland, J. Phys. A: Math. Gen. **23**, L621 (1990).
[38] H. Frauenkron, U. Bastolla, E. Gerstner, P. Grassberger and W. Nadler, Phys. Rev. Lett. **80**, 3149 (1998).
[39] J. Zhang, M. Lin, R. Chen, W. Wang and J. Liang, J. Chem. Phys. **128**, 125107 (2008).
[40] S. Cheluvaraja and H. Meirovitch, Proc. Natl. Acad. Sci. U.S.A. **101**, 9241 (2004).

[41]S. Cheluvaraja, M. Mihailescu and H. Meirovitch, J. Phys. Chem. B **112**, 9512 (2008).

[42]H.-P. Hsu and P. Grassberger, J Stat Phys **144**, 597 (2011).

[43]A. B. Mamonov, X. Zhang and D. M. Zuckerman, Journal of computational chemistry **32**, 396 (2011).

[44]J. P. Stoessel and P. Nowak, Macromolecules **23**, 1961 (1990).

[45]M. D. Tyka, A. R. Clarke and R. B. Sessions, J. Phys. Chem. B **110**, 17212 (2006).

[46]F. M. Ytreberg and D. M. Zuckerman, J. Chem. Phys. **124**, 104105 (2006).

[47]X. Zhang, A. B. Mamonov and D. M. Zuckerman, Journal of computational chemistry **30**, 1680 (2009).

[48]P. J. Flory, *Principles of Polymer Chemistry*. (Cornell University Press, Ithaca, 1953).

[49]P. G. de Gennes, *Scaling Concepts in Polymer Physics*. (Cornell University Press, Ithaca, 1979).

[50]A. I. U. Grosberg and A. R. Khokhlov, *Statistical Physics of Macromolecules*. (AIP Press, Woodbury, 1994).

[51]L. Schäfer, *Excluded volume effects in polymer solutions, as explained by the renormalization group*. (Springer, Berlin, 1999).

[52]K. Binder (Ed.), *Monte Carlo and Molecular Dynamics Simulations in Polymer Science*. (Oxford University Press, New York, 1995).

[53]M. Baiesi and E. Orlandini, Phys. Rev. E **86**, 031805 (2012).

[54]T. W. Burkhardt, Y. Yang and G. Gompper, Phys. Rev. E **82**, 041801 (2010).

[55]J. Gao, P. Tang, Y. Yang and J. Z. Y. Chen, Soft Matter **10**, 4674 (2014).

[56]N. G. Nossal, A. M. Makhov, P. D. Chastain 2nd, C. E. Jones and J. D. Griffith, J. Biol. Chem. **282**, 1098 (2007).

[57]M. J. Shoura, A. A. Vetcher, S. M. Giovan, F. Bardai, A. Bharadwaj, M. R. Kesinger and S. D. Levene, Nucleic Acids Res. **40**, 7452 (2012).

[58]J. Chaumeil, M. Micsinai, P. Ntziachristos, L. Deriano, J. M. Wang, Y. Ji, E. P. Nora, M. J. Rodesch, J. A. Jeddeloh, I. Aifantis, Y. Kluger, D. G. Schatz and J. A. Skok, Cell Rep. **3**, 359 (2013).

[59]Y. Jiang and P. E. Marszalek, EMBO J. **30**, 2881 (2011).

[60]N. Le May, D. Fradin, I. Iltis, P. Bougnères and J.-M. Egly, Mol. Cell **47**, 622 (2012).

[61]C. Hou, R. Dale and A. Dean, Proc. Natl. Acad. Sci. U.S.A. **107**, 3651 (2010).

[62]Y. Guo, K. Monahan, H. Wu, J. Gertz, K. E. Varley, W. Li, R. M. Myers, T. Maniatis and Q. Wu, Proc. Natl. Acad. Sci. U.S.A. **109**, 21081 (2012).

[63]A. Sanyal, B. R. Lajoie, G. Jain and J. Dekker, Nature **489**, 109 (2012).

[64]Z. Hensel, X. Weng, A. C. Lagda and J. Xiao, PLoS Biol. **11**, e1001591 (2013).

[65]A. V. Vologodskii, W. Zhang, V. V. Rybenkov, A. A. Podtelezhnikov, D. Subramanian, J. D. Griffith and N. R. Cozzarelli, Proc. Natl. Acad. Sci. U.S.A. **98**, 3045 (2001).

[66]K. Shimokawa, K. Ishihara, I. Grainge, D. J. Sherratt and M. Vazquez, Proc. Natl. Acad. Sci. U.S.A. **110**, 20906 (2013).

[67]N. J. Crisona, R. L. Weinberg, B. J. Peter, D. W. Sumners and N. R. Cozzarelli, J. Mol. Biol. **289**, 747 (1999).

[68]A. A. Vetcher, A. Y. Lushnikov, J. Navarra-Madsen, R. G. Scharein, Y. L. Lyubchenko, I. K. Darcy and S. D. Levene, J. Mol. Biol. **357**, 1089 (2006).

[69]A. Hanke and R. Metzler, Biophys. J. **85**, 167 (2003).

[70]M. A. Rubtsov, Y. S. Polikanov, V. A. Bondarenko, Y. H. Wang and V. M. Studitsky, Proc. Natl. Acad. Sci. U.S.A. **103**, 17690 (2006).

[71]Y. S. Polikanov, V. A. Bondarenko, V. Tchernaenko, Y. I. Jiang, L. C. Lutter, A. Vologodskii and V. M. Studitsky, Biophys. J. **93**, 2726 (2007).

[72]O. I. Kulaeva, E. V. Nizovtseva, Y. S. Polikanov, S. V. Ulianov and V. M. Studitsky, Mol. Cell Biol. **32**, 4892 (2012).

[73]K. V. Klenin and W. Wenzel, J. Chem. Phys. **139**, 054102 (2013).

[74]A. Gubaev and D. Klostermeier, Proc. Natl. Acad. Sci. U.S.A. **108**, 14085 (2011).

[75]N. Tokuda, M. Sasai and G. Chikenji, Biophys. J. **100**, 126 (2011).

[76]W. Li, D. Notani, Q. Ma, B. Tanasa, E. Nunez, A. Y. Chen, D. Merkurjev, J. Zhang, K. Ohgi, X. Song, S. Oh, H.-S. Kim, C. K. Glass and M. G. Rosenfeld, Nature **498**, 516 (2013).

[77]L. Yang, C. Lin, C. Jin, J. C. Yang, B. Tanasa, W. Li, D. Merkurjev, K. A. Ohgi, D. Meng, J. Zhang, C. P. Evans and M. G. Rosenfeld, Nature **500**, 598 (2013).

[78]D. T. Seaton, S. J. Mitchell and D. P. Landau, Braz. J. Phys. **36**, 623 (2006).

[79] J. Kierfeld, O. Niamploy, V. Sa-yakanit and R. Lipowsky, Eur. Phys. J. E Soft Matter **14**, 17 (2004).
[80] See Supplementary Material for mathematical and computational details and supporting figures; supporting movies are included in the version of this paper published in Journal of Chemical Physics.
[81] T. Biswas, H. Aihara, M. Radman-Livaja, D. Filman, A. Landy and T. Ellenberger, Nature **435**, 1059 (2005).
[82] Y. Chen, U. Narendra, L. E. Iype, M. M. Cox and P. A. Rice, Mol. Cell **6**, 885 (2000).
[83] D. N. Gopaul, F. Guo and G. D. Van Duyne, EMBO J. **17**, 4175 (1998).
[84] V. V. Rybenkov, N. R. Cozzarelli and A. V. Vologodskii, Proc. Natl. Acad. Sci. U.S.A. **90**, 5307 (1993).
[85] R. C. Tolman, *The Principles of Statistical Mechanics*. (Dover Publications, Inc., 2010).
[86] A. A. Podtelezhnikov, N. R. Cozzarelli and A. V. Vologodskii, Proc. Natl. Acad. Sci. U.S.A. **96**, 12974 (1999).
[87] G. Ziv, D. Thirumalai and G. Haran, Physical chemistry chemical physics : PCCP **11**, 83 (2009).
[88] Y. Burnier, J. Dorier and A. Stasiak, Nucleic Acids Res. **36**, 4956 (2008).
[89] J. D. Crawford, Reviews of Modern Physics **63**, 991 (1991).
[90] J.-U. Sommer, J. Chem. Phys. **97**, 5777 (1992).
[91] A. Hanke and R. Metzler, Chem. Phys. Lett. **359**, 22 (2002).
[92] S. G. Whittington and J. P. Valleau, J. Phys. A: Gen. Phys. **3**, 21 (1970).
[93] R. Metzler, A. Hanke, P. G. Dommersnes, Y. Kantor and M. Kardar, Phys. Rev. Lett. **88**, 188101 (2002).
[94] B. Marcone, E. Orlandini, A. L. Stella and F. Zonta, J. Phys. A: Math. Gen. **38**, L15 (2005).
[95] M. A. Cheston, K. McGregor, C. E. Soteros and M. L. Szafron, J. Stat. Mech. **2014**, P02014 (2014).

# Supplemental Material

## SI. Partition Function and Thermodynamics

We consider the conformational partition function of a semi-flexible harmonic chain freely floating in a volume $V$,

$$Q = \int_V \frac{d^3 r_1}{a^3} \cdots \int_V \frac{d^3 r_N}{a^3} \exp\left[-\beta U(\bar{r})\right] , \qquad (1)$$

where $\mathbf{r}_i$, $i = 1, \ldots, N$, are the positions of the $N$ vertices of the chain, $\beta = (k_B T)^{-1}$, and $U(\bar{r})$ is the total potential energy for a given chain conformation $\bar{r} = (\mathbf{r}_1, \ldots, \mathbf{r}_N)$ (cf. Section SII). The constant $a$ in Eq. (1) is a microscopic length required to make $Q$ dimensionless. For a system of massive point particles undergoing Newtonian dynamics, the length $a$ corresponds to the thermal wavelength; however, in this work we are only concerned with conformational degrees of freedom, and consider $a$ as a non-universal microscopic length much shorter than any other length scale associated with the chain. Essentially, the length $a$ corresponds to the lattice constant of an underlying lattice needed to obtain a finite number of accessible conformations. All results obtained in this work are independent of the length $a$, and thus largely independent of the discretization of the chain.

The mean energy is given by

$$\langle U \rangle = Q^{-1} \int_V \frac{d^3 r_1}{a^3} \cdots \int_V \frac{d^3 r_N}{a^3} U(\bar{r}) \exp\left[-\beta U(\bar{r})\right] . \qquad (2)$$

$\langle U \rangle$ is independent of the length $a$ in Eq. (1) because it drops out in the ratio in Eq. (2). Conversely, the *absolute* free energy

$$F = -k_B T \ln(Q) = \langle U \rangle - TS \qquad (3)$$

does depend on the length $a$ by the contribution from the entropy, $S$. However, in this work we are only concerned with the difference in free energy between, e.g., two knots $A$ and $B$ of equal length, which is given by

$$\Delta F_{AB} = F_B - F_A = -k_B T \ln(Q_B / Q_A) . \qquad (4)$$

The free-energy difference $\Delta F_{AB}$ is again independent of the length $a$ because it drops out in the ratio of partition functions in Eq. (4).

## SII. DNA-model Parameters

We consider a semi-flexible harmonic chain as a coarse-grained mesoscopic model for duplex DNA (cf. main text). Chain elements are extensible, cylindrical segments with equilibrium length $b_0$ and fixed diameter $d$, connected end-to-end by semi-flexible joints located at vertices $\mathbf{r}_i$, $i = 1,\ldots,N$. Segments are described by displacement vectors $\mathbf{b}_i = \mathbf{r}_{i+1} - \mathbf{r}_i$ with length $b_i$ and unit-length direction vectors $\hat{\mathbf{b}}_i = \mathbf{b}_i / b_i$. The total potential energy for a given conformation $\vec{r} = (\mathbf{r}_1,\ldots,\mathbf{r}_N)$ is defined as

$$U(\vec{r}) = U_{el} + U_{ev} + U_K, \tag{5}$$

where $U_{el}$ is the elastic energy of the chain (cf. Eq. (6) below) and $U_{ev}$, $U_K$ describe infinite potential barriers associated with excluded volume (overlapping chain segments) and changes in knot type $K$, respectively (Section SIII). These contributions are $U_{ev} = 0$ if none of the cylinders overlap and $U_{ev} = \infty$ otherwise. Similarly, $U_K = 0$ if the chain has knot type $K$ and $U_K = \infty$ otherwise.

The elastic energy for a conformation $\vec{r}$ is defined as

$$U_{el}(\vec{r}) = U_{sc} + k_B T \sum_{i=1}^{N} \left[ c_b \left(1 - \hat{\mathbf{b}}_i \cdot \hat{\mathbf{b}}_{i+1}\right) + \frac{c_s}{2}\left(\frac{b_i}{b_0} - 1\right)^2 \right], \tag{6}$$

where $c_b$, $c_s$ are bending and stretching elastic constants, respectively. The term $U_{sc}$ in Eq. (6) represents the elastic energy associated with deformations of the synaptic complex (sc) and does not contribute to the energy of unlooped circular chains. The location of the synaptic complex is determined by the number $n$ of segments in one of the loops. The contribution $U_{sc}$ is given by

$$U_{sc} = k_B T \frac{c_b}{2} \left( \hat{\mathbf{b}}_1 \cdot \hat{\mathbf{b}}_n + \hat{\mathbf{b}}_1 \cdot \hat{\mathbf{b}}_{n+1} + \hat{\mathbf{b}}_n \cdot \hat{\mathbf{b}}_N + \hat{\mathbf{b}}_{n+1} \cdot \hat{\mathbf{b}}_N \right). \tag{7}$$

The bending energy constant $c_b$ is chosen such that the persistence length of the chain, $P$, is equal to 5 segments.[1] Thus, $c_b$ is implicitly determined by the equation

$$\langle \cos(\theta) \rangle = \exp\left(-\frac{b_0}{P}\right) = \exp\left(-\frac{1}{5}\right) \tag{8}$$

where $\langle\cos(\theta)\rangle$ is a thermal average given by

$$\langle\cos(\theta)\rangle = \frac{\int_0^\pi d\theta \sin(\theta)\cos(\theta)\exp\left[-c_b(1-\cos(\theta))\right]}{\int_0^\pi d\theta \sin(\theta)\exp\left[-c_b(1-\cos(\theta))\right]} \quad . \tag{9}$$

Numerically solving Eq. (9) for $c_b$ yields $c_b = 5.5157$. We used $P = 50$ nm throughout this work,[2-4] corresponding to an equilibrium segment length of $b_0 = 10$ nm. The stretching energy constant $c_s$ is defined as $c_s = K_s b_0/(k_B T)$ where $K_s$ is the stretch modulus. Using $b_0 = 10$ nm, $T = 300$ K, and the approximate value $K_s = 1000$ pN for DNA under physiological conditions,[5] we obtain $c_s \approx 2500$ which is used throughout this work. Excluded-volume and electrostatic interactions between DNA segments are modeled by an effective hard-cylinder diameter, $d$. We used $d = 0.5 b_0 = 5$ nm throughout this work, corresponding to an ionic strength of 150 mM.[6]

### SIII. Monte Carlo Simulation Procedure

To calculate averages such as $\langle U \rangle$ in Eq. (2) we used a Metropolis Monte Carlo procedure to generate equilibrium ensembles of chain conformations. The procedure builds on previous calculations described in [1]. Trial conformations were accepted with probability $P_{accept}$ according to the Metropolis criterion, namely

$$P_{accept} = \min\left[\exp\left[-\beta(U_{trial} - U_{current})\right], 1\right] \tag{10}$$

where $U_{trial}$ and $U_{current}$ are the potential energies of trial and current conformations, respectively (Eq. (5)).

Trial moves applied to knotted, circular chains were crankshaft rotations and segment stretching moves (Fig. S1A, B). For both types of trial moves, two randomly selected vertices $\mathbf{r}_i$ and $\mathbf{r}_j$ define the subsection of the chain to be displaced. Crankshaft moves rotate the chosen subsection by an angle $\psi$ about the axis $\mathbf{r}_j - \mathbf{r}_i$. The rotation angle $\psi$ is chosen uniformly from an interval $[-\psi_{max}, \psi_{max}]$. Stretching moves translate the chosen subsection by a displacement vector $\mathbf{s}$ with random magnitude and direction. The magnitude of $\mathbf{s}$ is chosen uniformly from an interval $[0, s_{max}]$. The parameters $s_{max}$ and $\psi_{max}$ are dynamically updated so that ~50% of both types of trial moves is accepted.

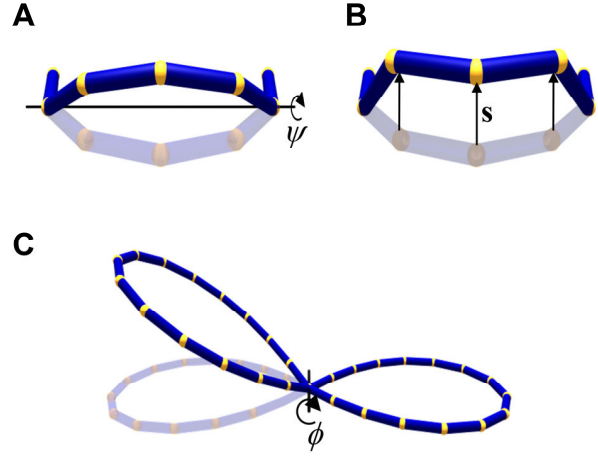

FIG. S1. Trial moves applied during Monte Carlo simulations. (A) Crankshaft moves rotate a section of the chain between vertices $\mathbf{r}_i$ and $\mathbf{r}_j$ about an axis $\mathbf{r}_j - \mathbf{r}_i$ (indicated by the black bar) by an angle $\psi$. (B) Stretching moves displace all vertices between $\mathbf{r}_i$ and $\mathbf{r}_j$ by the same displacement vector $\mathbf{s}$. (C) Additional trial moves applied to dual-loop sc models rotate one loop about a randomly oriented axis through $\mathbf{r}_1$ (indicated by the black bar) by an angle $\phi$.

For dual-loop synaptic complex (sc) models, crankshaft and stretching trial moves as described above were confined to single-looped domains in order to preserve the constraint $\mathbf{r}_1 \equiv \mathbf{r}_{n+1} \equiv \mathbf{r}_{N+1}$. The corresponding loop is randomly selected and the vertices $\mathbf{r}_i$, $\mathbf{r}_j$ for these trial moves are selected randomly within the chosen loop. The common vertex $\mathbf{r}_1$ of the two loops is never displaced. An additional type of trial move was applied specifically to dual-loop sc models (Fig. S1C). The additional trial move rotates one of the loops by an angle $\phi$ about a randomly oriented axis through $\mathbf{r}_1$. The rotation angle $\phi$ is chosen uniformly from an interval $[-\phi_{max}, \phi_{max}]$ where the parameter $\phi_{max}$ is again dynamically updated to accept ~ 50% of these trial moves.

A single trial move was applied at each iteration of the Monte Carlo simulation. The type of trial move was chosen randomly. Crankshaft and stretching moves were attempted with equal probability. For dual-loop sc models, loop-rotation trial moves were attempted with 2% probability.

After applying a trial move to the current conformation, the Metropolis criterion in Eq. (10) was evaluated using the elastic energies of trial and current conformations. If the trial move was accepted based on comparing elastic energies alone, $U_{ev}$ and $U_K$ were enforced to ensure that none of the chain segments overlapped and that knot type $K$ was conserved. To enforce $U_{ev}$, the distance of closest approach for all pairs of non-adjacent cylinders was computed and compared with the cylinder diameter $d$ to ensure that no cylinders overlap. The trial conformation was rejected if the distance between any pairs of segments was found to be less than $d$. To enforce $U_K$, knotted chains were verified to having correct knot type $K$ by computing the HOMFLY polynomial. In previous calculations, the Alexander polynomial, $\Delta_K(t)$, was used to verify knot type of chains generated during Monte Carlo simulation.[1,6,7] However, $\Delta_K(t)$ cannot distinguish any chiral knot, $K$, from its topologically different mirror image, $K^*$. We therefore used the more powerful HOMFLY polynomial, $\rho_K(l,m)$, to uniquely identify the knot type.[8] The HOMFLY polynomial identifies knots, including their chirality, with up to 10 irreducible crossings except for the knots $9_{42}$ and $10_{71}$.[9] HOMFLY calculations can potentially be rate limiting especially for knots with many crossings. For the chain lengths and knot types modeled here, efficiency of the HOMFLY calculation was significantly improved by choosing an orientation (projection) which produces the fewest possible crossings.

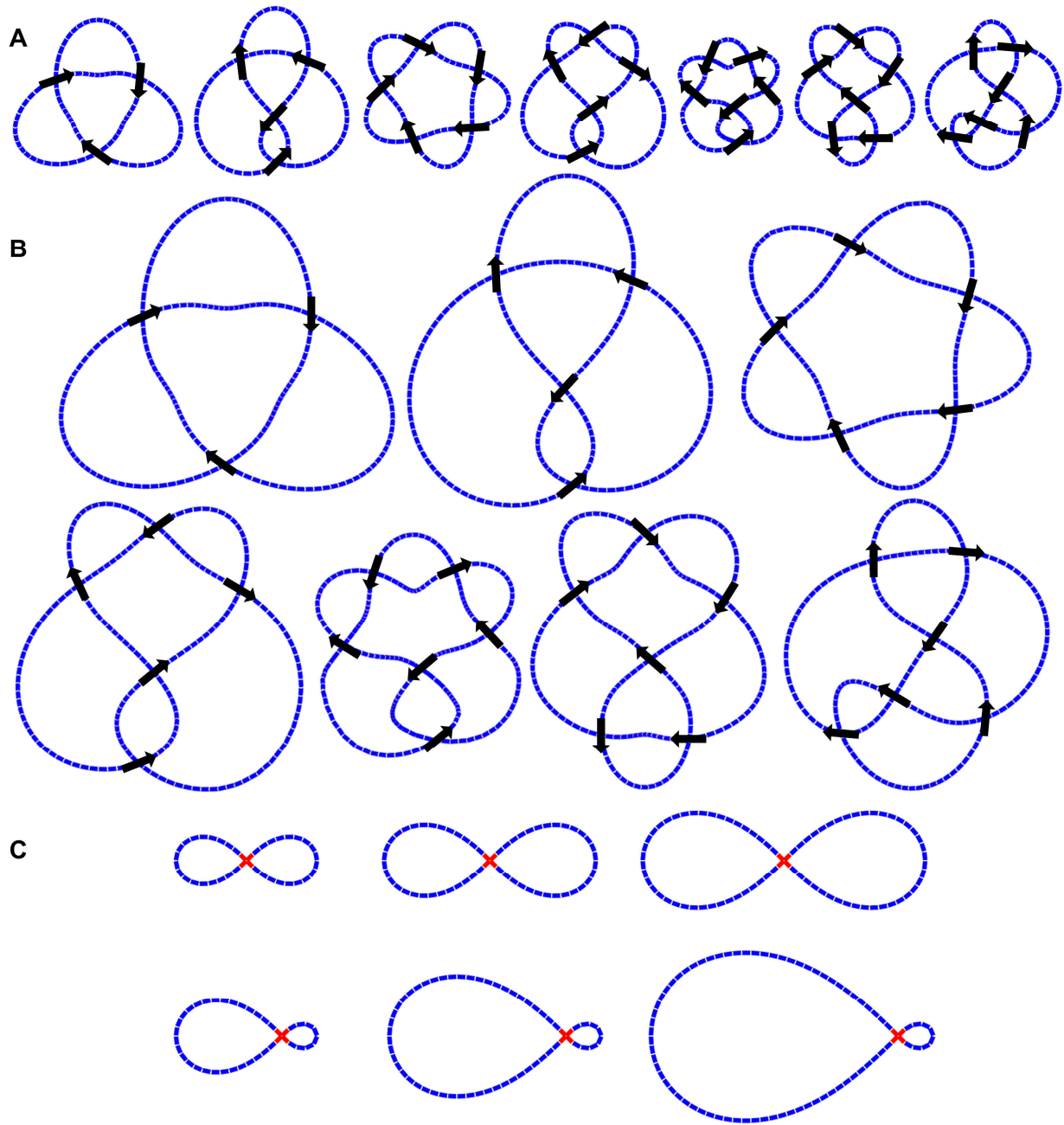

FIG. S2. Reference conformations (to scale). (A) Reference conformations of knotted chains with $N = 100$ segments ($L = 20P$, representing DNA molecules with 3000 base pairs). Black arrows indicate the direction of the over-crossing segment. (B) Reference conformations of knotted chains with $N = 200$ ($L = 40P$, 6000 base pairs). (C) Examples of reference conformations used for dual-loop sc models. Segments shown in red represent the location of the synapse. Left column: chains with $N = 40$ segments ($L = 8P$, 1200 base pairs); middle column: $N = 60$ segments ($L = 12P$, 1800 base pairs); right column: $N = 80$ segments ($L = 16P$, 2400 base pairs). For all chains shown in the top row the location of the synapse is midway along the contour of the molecule, i.e., $\ell = L/2$. The smaller loops shown in the bottom row all have the same length $\ell = 2P$ ($n = 10$).

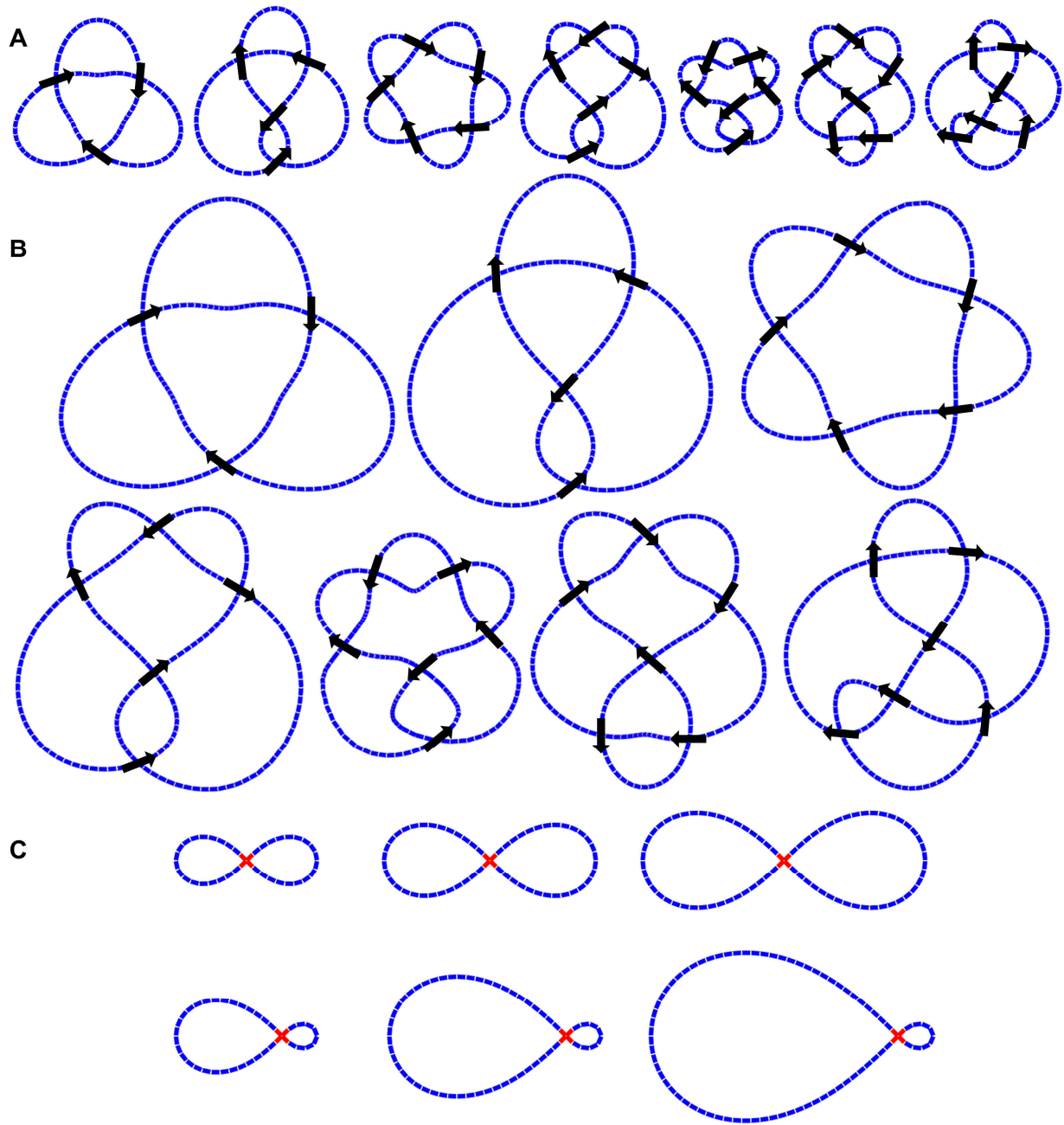

FIG. S2. Reference conformations (to scale). (A) Reference conformations of knotted chains with $N = 100$ segments ($L = 20P$, representing DNA molecules with 3000 base pairs). Black arrows indicate the direction of the over-crossing segment. (B) Reference conformations of knotted chains with $N = 200$ ($L = 40P$, 6000 base pairs). (C) Examples of reference conformations used for dual-loop sc models. Segments shown in red represent the location of the synapse. Left column: chains with $N = 40$ segments ($L = 8P$, 1200 base pairs); middle column: $N = 60$ segments ($L = 12P$, 1800 base pairs); right column: $N = 80$ segments ($L = 16P$, 2400 base pairs). For all chains shown in the top row the location of the synapse is midway along the contour of the molecule, i.e., $\ell = L/2$. The smaller loops shown in the bottom row all have the same length $\ell = 2P$ ($n = 10$).

Monte Carlo simulations were started with one of the reference conformations shown in Fig. S2. Knotted ground-state conformations $\vec{r}_0$ were obtained using dynamics in KnotPlot[10] by adjusting segment lengths and adding a weak self-repulsive potential (Fig. S2A, B). We found that this choice of minimum seems to yield the most efficient and precise TI-NMA results (although as discussed in section III.A.2 of the main text, the computed free energy does not measurably depend on the choice of minimum). Ground state conformations $\vec{r}_0$ for synaptic complex models (Fig. S2C) were obtained by BFGS energy minimization of $U_{el}(\vec{r})$ (Eq. (6)) as implemented in MATLAB.[11]

### SIV. Equilibrium Segment Passage

The ratio of partition functions $Q_B/Q_A$ in Eq. (4) is equal to the probability ratio $P_B/P_A$ in any equilibrium ensemble in which the two knots $A$ and $B$ coexist, where $P_A$ is the probability of observing $A$ and $P_B$ is the probability of observing $B$. In a Metropolis Monte Carlo simulation (Section SIII) such an ensemble is obtained by allowing changes in knot type during successive deformations of the chain. For example, by choosing $U_K = 0$ in Eq. (5) of the main text for knot types $A$ or $B$ but $U_K = \infty$ otherwise, these knots will coexist in an equilibrium ensemble if there exists trial moves which transform $A$ into $B$ and vice versa. Using $Q_B/Q_A = P_B/P_A$ in Eq. (4) one finds that the free energy difference $\Delta F_{AB}$ can be directly inferred from $P_B/P_A$ in the Equilibrium Strand Passage (ESP) ensemble,[7] i.e.,

$$\Delta F_{AB} = -k_B T \ln(P_B/P_A) \,. \tag{11}$$

Equation (11) holds for any ensemble in which $A$, $B$ coexist in equilibrium, independent of which other knot types are present in the ensemble. Consider, for example, an ESP ensemble in which three knot types, $A$, $B$ and $C$ coexist and which is dominated by $C$ such that $P_C \gg P_A, P_B$. Obtaining $\Delta F_{AB}$ accurately in this ensemble by using Eq. (11) is numerically challenging because $A$, $B$ are observed only with low frequency. It is therefore advantageous to exclude the dominating knot type $C$ from the ensemble, so that $A$ and $B$ are observed with higher frequency, and to obtain $\Delta F_{AB}$ using Eq. (11) in the restricted ensemble. In our calculations we consider ensembles in which no knot types are restricted, ESP-$\{\varnothing\}$, as well as the restricted ensembles ESP-$\{0_1\}$ and ESP-$\{0_1, 3_1, 3_1^*\}$ to accurately obtain relative probabilities of knots having large free energy (Table S1).

TABLE SI. Knotting probabilities obtained by Equilibrium Strand Passage (ESP)

| Knot-Type | $L/P = 20$, $N = 100$ | | | $L/P = 40$, $N = 200$ | | |
|---|---|---|---|---|---|---|
| | ESP-$\{\varnothing\}$ | ESP-$\{0_1\}$ | ESP-$\{0_1, 3_1, 3_1^*\}$ | ESP-$\{\varnothing\}$ | ESP-$\{0_1\}$ | ESP-$\{0_1, 3_1, 3_1^*\}$ |
| $0_1$ | 997.95(0.02)[a] | - | - | 989.36(0.07) | - | - |
| $3_1$ | 0.99(0.02) | 483(15) | - | 4.9(0.1) | 462(2) | - |
| $3_1^*$ | 0.98(0.01) | 480(15) | - | 4.9(0.1) | 453(3) | - |
| $4_1$ | 0.07(0.01) | 32(2) | 846(4) | 0.62(0.02) | 62(1) | 713(2) |
| $5_1$ | | 0.87(0.02) | 35(1) | 0.04(0.01) | 3.9(0.1) | 50(1) |
| $5_1^*$ | | 0.92(0.05) | 36(1) | 0.04(0.01) | 3.8(0.1) | 50(1) |
| $5_2$ | | 0.10(0.02) | 38(1) | 0.06(0.01) | 5.4(0.1) | 69(1) |
| $5_2^*$ | | 0.10(0.03) | 39(1) | 0.06(0.01) | 5.3(0.1) | 70(1) |
| $6_1$ | | 0.03(0.01) | 0.76(0.01) | | 0.40(0.01) | 4.6(0.1) |
| $6_1^*$ | | 0.03(0.01) | 0.77(0.01) | | 0.42(0.02) | 4.7(0.1) |
| $6_2$ | | 0.02(0.01) | 0.71(0.02) | | 0.39(0.01) | 4.5(0.1) |
| $6_2^*$ | | 0.02(0.01) | 0.66(0.03) | | 0.39(0.01) | 4.5(0.1) |
| $6_3$ | | 0.02(0.01) | 0.91(0.01) | | 0.35(0.01) | 5.3(0.1) |
| Other | | 0.07(0.01) | 2.6(0.1) | | 1.9(0.1) | 25(1) |

[a]Expected number of knots per 1000 conformations. Values in parentheses represent SEM (10 trials, each with $\eta = 5 \times 10^5$ conformations)

**SV. Thermodynamic Integration**

The objective of thermodynamic integration (TI) is to replace the energy function of the original, semi-flexible system described by Eq. (5) with a potential function suitable for NMA, and to calculate the associated change in free energy by TI (Fig. 4). To this end we define

$$U(\lambda) = \begin{cases} \lambda U_{ha} + (1-\lambda) U_{el} + U_{ev} + U_K, & 0 \leq \lambda \leq 1 \\ \lambda U_{ha} + U_{ev} + U_K, & 1 \leq \lambda \leq \lambda_{max} \end{cases}, \quad (12)$$

where the auxiliary elastic energy $U_{ha}$ serves to constrain the system to the predefined reference conformation. The term $U_{ha}$ is given by (main text, Eq. (8))

$$U_{ha}(\vec{r}, \vec{r}_0) = U_{sc} + k_B T \sum_{i=1}^{N} \left[ \frac{c_b}{2} \left( 2 - \hat{\mathbf{b}}_{i+1} \cdot \hat{\mathbf{b}}_{i+1}^p - \hat{\mathbf{b}}_{i-1} \cdot \hat{\mathbf{b}}_{i-1}^p \right) + \frac{c_s}{2} \left( \frac{b_i}{b_0} - 1 \right)^2 \right]. \quad (13)$$

TI is carried out in two phases given by (main text, Eqs. [15], [16])

$$\Delta F_1 = \int_0^1 d\lambda \left\langle \frac{dU}{d\lambda} \right\rangle_\lambda = \int_0^1 d\lambda \left\langle U_{ha} - U_{el} \right\rangle_\lambda, \tag{14}$$

$$\Delta F_2 = \int_1^{\lambda_{max}} d\lambda \left\langle \frac{dU}{d\lambda} \right\rangle_\lambda = \int_1^{\lambda_{max}} d\lambda \left\langle U_{ha} \right\rangle_\lambda, \tag{15}$$

where $U = U(\lambda)$ (Eq. (12)). Values for $\left\langle U_{ha} - U_{el} \right\rangle_\lambda$ in Eq. (14) and $\left\langle U_{ha} \right\rangle_\lambda$ in Eq. (15) were obtained by Monte Carlo simulation for 21 equally spaced values of $\lambda = \{0, 0.05, 0.1, \ldots, 1\}$ and for 101 exponentially increasing values from 1 to $\lambda_{max} = 200$. The results were then linearly interpolated and integrated according to Eqs. (14), (15). Each simulation was started at a reference conformation (Fig. S2), and an initial $2 \times 10^6$ trial moves were made to equilibrate the system. Following equilibration, a new conformation was saved after each 2000 trial moves to produce an ensemble of $\eta$ independent conformations. Figure S3 shows $\left\langle U_{ha} - U_{el} \right\rangle_\lambda$ and $\left\langle \lambda U_{ha} \right\rangle_\lambda$ for all values of $\lambda$ quoted above and for all knotted configurations studied in this work. Each data point represents the average obtained from an ensemble of $\eta = 5 \times 10^5$ conformations ($\eta = 10^6$ in the case of the unknot, $0_1$). Error (SEM) in the measurement was estimated by dividing each ensemble into 5 smaller ensembles and calculating standard error between these smaller ensembles.

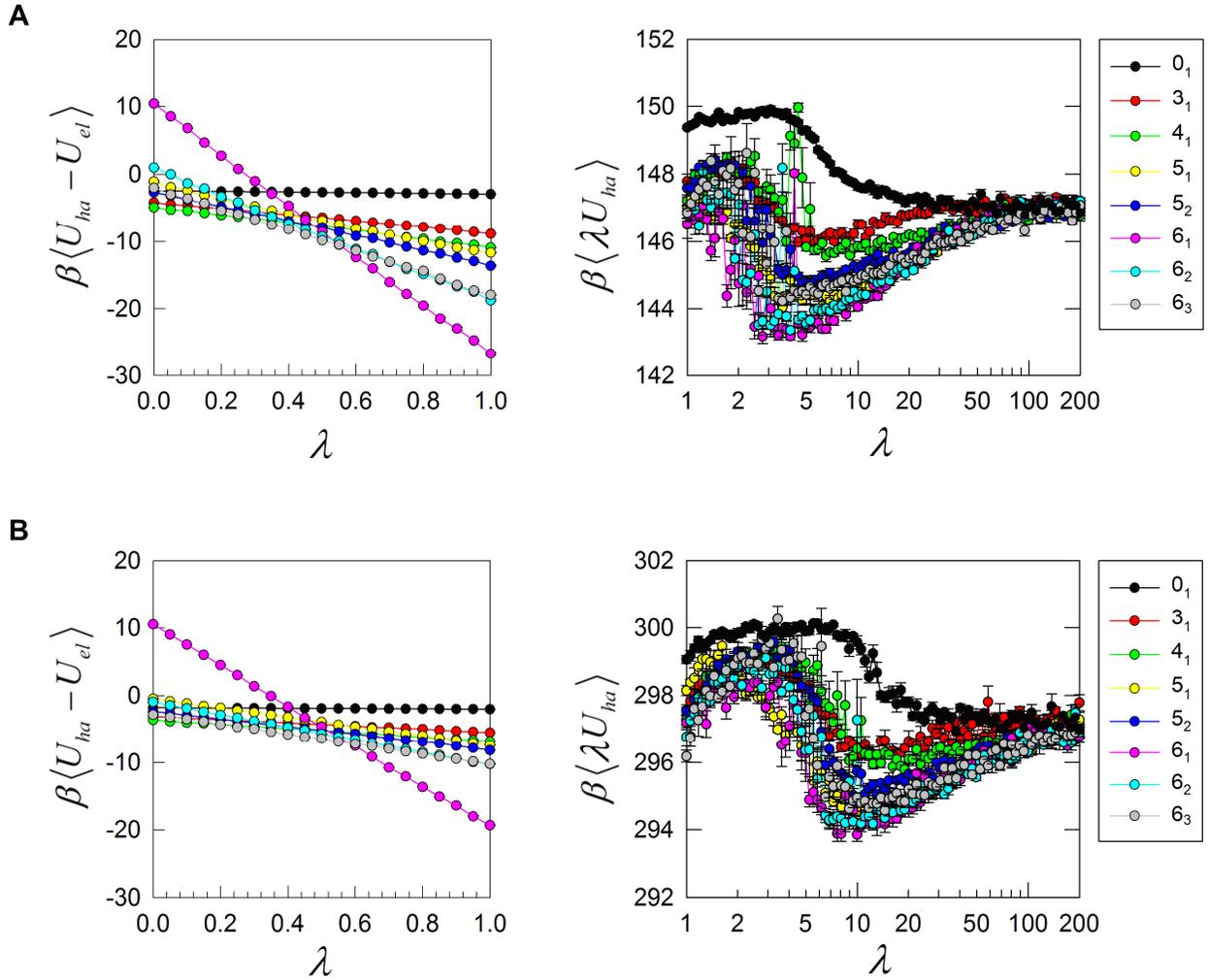

FIG. S3. Values of $\langle U_{ha} - U_{el} \rangle_\lambda$ (left) and $\langle \lambda U_{ha} \rangle_\lambda$ (right) obtained for knotted chains during thermodynamic integration. Graphs shown in the top and bottom row are for chains with $N = 100$ and $N = 200$ segments, respectively. $\beta \langle \lambda U_{ha} \rangle_\lambda$ converges for large $\lambda$ to $(3N-6)/2 = (147, 297)$ for $N = (100, 200)$ according to the equipartition theorem (cf. Eq. (20)).

**SVI. Normal-mode Analysis**

In what follows we derive the free energy $F_0$ of the minimum-energy (ground state) conformation from Normal-mode Analysis (NMA) (main text, Eq. [17]). We consider the potential energy $U(\vec{r})$ of a reference system governed by harmonic vibrations about a minimum-energy conformation $\vec{r}_0$. To simplify notation we combine position vectors $\mathbf{r}_i = (x_{i,1}, x_{i,2}, x_{i,3})$ in a single vector $\vec{r} = (\mathbf{r}_1, ..., \mathbf{r}_N)$ of $3N$ components $x_{i,\nu}$ indexed by $i = 1, ..., N$, $\nu = 1, 2, 3$. NMA is implemented by expanding $U(\vec{r})$ about $\vec{r}_0$, i.e.,

$$\beta U(\vec{r}) = \beta E_0 + \frac{1}{2b_0^2}(\vec{r} - \vec{r}_0) \cdot \mathbf{H}(\vec{r} - \vec{r}_0) + ...$$
$$= \beta E_0 + \frac{1}{2b_0^2} \sum_{m=1}^{3N} \nu_m q_m^2 + ... \qquad (16)$$

and disregarding terms of higher than second order in the deviation $\vec{r} - \vec{r}_0$, as indicated by the dots in Eq. (16). $E_0 = U(\vec{r}_0)$ is the minimum energy (chosen to be $E_0 = 0$ by construction) and we have used $\vec{\nabla} U(\vec{r}_0) = 0$. The dot symbol in Eq. (16) denotes the usual dot product and $\mathbf{H}$ is the $(3N) \times (3N)$ Hessian matrix with elements

$$H_{i\mu, j\nu} = \beta b_0^2 \frac{\partial^2 U(\vec{r}_0)}{\partial x_{i,\mu} \partial x_{j,\nu}}, \quad i, j = 1, ..., N, \quad \mu, \nu = 1, 2, 3. \qquad (17)$$

$\mathbf{H}$ is dimensionless as energies are expressed in units of $k_B T = \beta^{-1}$ and lengths in units of the average segment length $b_0$. In the second line of Eq. (16), $\nu_m$ are the $3N$ (unitless) eigenvalues of $\mathbf{H}$ and $q_m$ are the corresponding normal modes.

## A. Partition function and free energy

Upon inserting Eq. (16) in Eq. (1) and considering degrees of freedom associated with rigid translations and rotations of the minimum conformation $\vec{r}_0$ in the volume $V$ we obtain the NMA expression of the partition function

$$Q_{NMA} = \exp(-\beta E_0) \frac{V}{a^3} \left(\frac{b_0^3}{a^3}\right)^{N-1} N^{3/2} 8\pi^2 \sqrt{I_x I_y I_z} \prod_{m=7}^{3N} \left(\frac{2\pi}{\nu_m}\right)^{1/2}. \qquad (18)$$

$I_x, I_y, I_z$ in Eq. (18) are the principal moments of inertia[a] of the minimum-energy conformation $\vec{r}_0$ in units of $b_0$. The free energy in units of $k_B T$ obtained in NMA is given by

$$\beta F_0 = -\ln(Q_{NMA})$$
$$= \beta E_0 - \rho - \ln\left(N^{3/2} 8\pi^2 \sqrt{I_x I_y I_z}\right) - \frac{1}{2}\sum_{m=7}^{3N} \ln\left(\frac{2\pi}{\nu_m}\right), \quad (19)$$

where $\rho = \ln\left[\frac{V}{a^3}\left(\frac{b_0^3}{a^3}\right)^{N-1}\right]$. The value of $\rho$ depends on the discretization of the system but drops out in differences taken between systems with the same number of segments $N$. The mean energy in units of $k_B T$ obtained in NMA is given by (as expected from the equipartition theorem)

$$\langle \beta U_0 \rangle = -\beta \frac{\partial}{\partial \beta} \ln(Q_{NMA}) = \beta E_0 + \frac{1}{2}(3N - 6). \quad (20)$$

---

[a] We here assume that each vertex of the chain at position $\mathbf{r}_i$ is associated with a unit mass.

Here we have used that $\nu_m$ in Eq. (19) depends on $\beta$ in the form $\nu_m = \beta c_m$ where $c_m$ is an effective spring constant associated with the normal mode $m$.

### B. Normal modes

Assuming that the eigenvalues $\nu_m$ of $\mathbf{H}$ are ordered such that $\nu_1 \leq \nu_2 \leq \ldots \leq \nu_{3N}$ one finds $\nu_m = 0$ for $m = 1, \ldots, 6$ and $\nu_m > 0$ for $m = 7, \ldots, 3N$. The $3N - 6$ nonzero eigenvalues $\nu_m > 0$ for $m = 7, \ldots, 3N$ are associated with internal vibrations of the chain about the minimum conformation $\vec{r}_0$ which incur a finite energetic cost. Conversely, the 6 zero eigenvalues $\nu_m = 0$ for $m = 1, \ldots, 6$ are associated with rigid translations and rotations of the system which do not incur any energetic cost. The corresponding modes $q_m$, $m = 1, \ldots, 6$, contribute to $Q_{NMA}$ in Eq. (18) in terms of the number $N$ of "particles" (joints) and on the shape of the minimum conformation, $\vec{r}_0$, in terms of $I_x, I_y, I_z$. The dependence of the entropy on total mass and moments of inertia for systems governed by Newtonian dynamics is usually ignored; see [12], page 12661 and K. Hinsen in [13], page 7.

In our calculations, the minimum conformation $\vec{r}_0$ is translated and rotated such that the center of mass is at $(0,0,0)$ and the inertia tensor is diagonal in Cartesian coordinates. Principal moments of inertia $I_x, I_y, I_z$ of the minimum conformation were calculated as eigenvalues of the inertia tensor, given by the diagonal elements in the principal-axes frame. The Hessian matrix $\mathbf{H}$ in Eq. (17) is obtained by calculating $U(\vec{r})$ for values of $\vec{r} = (\mathbf{r}_1,...,\mathbf{r}_N)$ in a region about $\vec{r}_0$ on a 5x5 grid of grid size $a = 0.01$ (in units of $b_0$). First and second partial derivatives in Eq. (17) were calculated by cubic spline interpolation between grid points. Eigenvalues $v_m$ of $\mathbf{H}$ are calculated by Cholesky factorization within MATLAB. We verified in all cases that 6 eigenvalues were zero to numerical precision, and that the remaining $3N - 6$ eigenvalues were positive.

Because contributions from rigid translations and rotations are usually neglected, but nevertheless important, in what follows we derive Eqs. (18) and (19) in detail. The normal modes $q_m$ in Eq. (16) are given by

$$q_m = \hat{u}^{(m)} \cdot (\vec{r} - \vec{r}_0) = \sum_{i=1}^{N} \mathbf{u}_i^{(m)} \cdot \left(\mathbf{r}_i - \mathbf{r}_i^{(0)}\right), \quad m = 1,...,3N, \tag{21}$$

where $\hat{u}^{(m)} = \left(\mathbf{u}_1^{(m)},...,\mathbf{u}_N^{(m)}\right)$ is a normalized eigenvector of $\mathbf{H}$ with eigenvalue $v_m$, i.e., $\mathbf{H}\hat{u}^{(m)} = v_m \hat{u}^{(m)}$ (the hat symbol indicates that $\hat{u}^{(m)}$ is normalized). Since $\mathbf{H}$ is symmetric the eigenvectors $\hat{u}^{(m)}$ are orthonormal. Thus

$$\hat{u}^{(m)} \cdot \hat{u}^{(n)} = \sum_{i=1}^{N} \mathbf{u}_i^{(m)} \cdot \mathbf{u}_i^{(n)} = \delta_{mn}, \quad m,n = 1,...,3N. \tag{22}$$

The eigenvectors $\hat{u}^{(m)}$ for $m = 7,...,3N$ with positive eigenvalues $v_m > 0$ are associated with internal vibrations of the system which incur a finite energetic cost. Upon inserting Eq. (21) in Eq. (18) one finds

$$\begin{aligned} Q_{NMA} &= \exp(-\beta E_0) \, a^{-3N} J \left( \prod_{m=1}^{6} \int dq_m \right) \prod_{m=7}^{3N} \int_{-\infty}^{\infty} dq_m \exp\left[-\frac{1}{2b_0^2} \sum_{m=7}^{3N} v_m q_m^2 \right] \\ &= \exp(-\beta E_0) \, a^{-3N} \left( \prod_{m=1}^{6} \int dq_m \right) b_0^{3N-6} \prod_{m=7}^{3N} \left(\frac{2\pi}{v_m}\right)^{1/2} \end{aligned} \tag{23}$$

The factor $J$ in the first line is the Jacobian factor for the transformation from Cartesian coordinates $\mathbf{r}_i$ in Eq. (1) to normal modes $q_m$ in Eq. (18). This transformation is orthogonal, hence $J = 1$. The term $\prod_{m=1}^{6} \int dq_m$ in Eq. (23) is associated with rigid translations and rotations of the system and will be discussed in the following section.

## C. Rigid translations and rotations

The set $\{\hat{u}^{(m)}, m=1,\ldots,3N\}$ includes 3 eigenvectors $\hat{u}^{(1)}$, $\hat{u}^{(2)}$, $\hat{u}^{(3)}$ associated with rigid translations and 3 eigenvectors $\hat{u}^{(4)}$, $\hat{u}^{(5)}$, $\hat{u}^{(6)}$ associated with rigid rotations of the minimum conformation $\vec{r}_0$. For the translational eigenvectors we choose

$$\hat{u}^{(v)} = \frac{1}{\sqrt{N}}\left(\mathbf{u}_1^{(v)},\ldots,\mathbf{u}_N^{(v)}\right), \quad v=1,2,3, \tag{24}$$

with

$$\mathbf{u}_i^{(1)} = (1,0,0), \quad \mathbf{u}_i^{(2)} = (0,1,0), \quad \mathbf{u}_i^{(3)} = (0,0,1), \quad i=1,\ldots,N. \tag{25}$$

The factor $1/\sqrt{N}$ in (24) arises from the normalization condition $\hat{u}^{(v)} \cdot \hat{u}^{(v)} = 1$. Using (24) one obtains $q_v = \frac{1}{\sqrt{N}}\sum_{i=1}^{N}\left(x_{i,v} - x_{i,v}^{(0)}\right)$, thus $dq_v = \frac{1}{\sqrt{N}}\sum_{i=1}^{N}dx_{i,v}$, $v=1,2,3$. In terms of the mass center, $\mathbf{R}_{cm} = \frac{1}{N}\sum_{i=1}^{N}\mathbf{r}_i$, one finds $dq_v = \sqrt{N}dR_{cm,v}$. The translational degrees of freedom of the system thus amount to an $N$-dependent contribution

$$\int dq_1 \int dq_2 \int dq_3 = N^{3/2}\int_V d^3 R_{cm} = VN^{3/2}. \tag{26}$$

The rotational eigenvectors $\hat{u}^{(4)}$, $\hat{u}^{(5)}$, $\hat{u}^{(6)}$ should be chosen such that the orthonormality condition (22) holds for $m,n=1,\ldots,6$. For given minimum conformation $\vec{r}_0$ this condition uniquely specifies the rotational eigenvectors as follows: (i) The axes of rotation associated with $\hat{u}^{(4)}$, $\hat{u}^{(5)}$, $\hat{u}^{(6)}$ coincide with the principal axes of $\vec{r}_0$. This implies that the axes of rotation pass through the center of mass[a] of $\vec{r}_0$ given by $\mathbf{R}_{cm}^{(0)} = \frac{1}{N}\sum_{i=1}^{N}\mathbf{r}_i^{(0)}$. Without restriction, the coordinate frame $(\hat{\mathbf{e}}_x, \hat{\mathbf{e}}_y, \hat{\mathbf{e}}_z)$ may be chosen to be centered at $\mathbf{R}_{cm}^{(0)}$, so that $\mathbf{R}_{cm}^{(0)} = 0$ relative to $(\hat{\mathbf{e}}_x, \hat{\mathbf{e}}_y, \hat{\mathbf{e}}_z)$, and such that it coincides with the principal axes of $\vec{r}_0$. This is assumed to be the case in what follows. (ii) Under the assumptions in (i), the rotational eigenvectors are then given by

$$\hat{u}^{(v)} = k_v\left(\mathbf{u}_1^{(v)},\ldots,\mathbf{u}_N^{(v)}\right), \quad v=4,5,6, \tag{27}$$

where $k_\nu$ are normalization constants so that $\hat{u}^{(\nu)} \cdot \hat{u}^{(\nu)} = 1$ and

$$\mathbf{u}_i^{(4)} = J_x \mathbf{r}_i^{(0)} \quad, \quad \mathbf{u}_i^{(5)} = J_y \mathbf{r}_i^{(0)} \quad, \quad \mathbf{u}_i^{(6)} = J_z \mathbf{r}_i^{(0)} \quad, \quad i = 1, \ldots, N \quad. \tag{28}$$

Here we use generators of infinitesimal rotations about $\hat{\mathbf{e}}_x$, $\hat{\mathbf{e}}_y$, $\hat{\mathbf{e}}_z$ given by [14]

$$J_x = \begin{pmatrix} 0 & 0 & 0 \\ 0 & 0 & -1 \\ 0 & 1 & 0 \end{pmatrix}, \quad J_y = \begin{pmatrix} 0 & 0 & 1 \\ 0 & 0 & 0 \\ -1 & 0 & 0 \end{pmatrix}, \quad J_z = \begin{pmatrix} 0 & -1 & 0 \\ 1 & 0 & 0 \\ 0 & 0 & 0 \end{pmatrix}. \tag{29}$$

Note that, according to Eq. (28), the rotational eigenvectors $\hat{u}^{(4)}$, $\hat{u}^{(5)}$, $\hat{u}^{(6)}$ depend on the minimum conformation $\vec{r}_0$. One readily verifies condition (22) for $\hat{u}^{(1)}, \ldots, \hat{u}^{(6)}$. For example,

$$\hat{u}^{(1)} \cdot \hat{u}^{(5)} = \frac{k_5}{\sqrt{N}} \sum_{i=1}^{N} (1,0,0) \cdot \left( z_i^{(0)}, 0, -x_i^{(0)} \right) = \frac{k_5}{\sqrt{N}} \sum_{i=1}^{N} z_i^{(0)} = 0 \quad \text{since} \quad \mathbf{R}_{cm}^{(0)} = 0. \quad \text{Similarly,}$$

$\hat{u}^{(4)} \cdot \hat{u}^{(5)} = k_4 k_5 \sum_{i=1}^{N} \left( 0, -z_i^{(0)}, y_i^{(0)} \right) \cdot \left( z_i^{(0)}, 0, -x_i^{(0)} \right) = -k_4 k_5 \sum_{i=1}^{N} x_i^{(0)} y_i^{(0)} = 0$; this follows because the inertia tensor of $\vec{r}_0$,

$$\Gamma_{\mu\nu} = \sum_{i=1}^{N} \left( \delta_{\mu\nu} \mathbf{r}_i^{(0)} \cdot \mathbf{r}_i^{(0)} - x_{i\mu}^{(0)} x_{i\nu}^{(0)} \right) \quad, \tag{30}$$

is diagonal in the principal-axis frame by construction, hence $\sum_{i=1}^{N} x_i^{(0)} y_i^{(0)} = \Gamma_{12} = 0$. Using the condition $\hat{u}^{(\nu)} \cdot \hat{u}^{(\nu)} = 1$ for $\nu = 4, 5, 6$ one obtains the normalization constants $k_\nu$ in (27) as

$$k_4 = \left( b_0^2 I_x \right)^{-1/2}, \quad k_5 = \left( b_0^2 I_y \right)^{-1/2}, \quad k_6 = \left( b_0^2 I_z \right)^{-1/2} \tag{31}$$

where $I_\nu = \Gamma_{\nu\nu} / b_0^2$ are the principal moments of inertia[a] of $\vec{r}_0$ in units of $b_0$. For example,

$1 = \hat{u}^{(4)} \cdot \hat{u}^{(4)} = k_4^2 \sum_{i=1}^{N} \left( 0, -z_i^{(0)}, y_i^{(0)} \right)^2 = k_4^2 \sum_{i=1}^{N} \left[ \left( z_i^{(0)} \right)^2 + \left( y_i^{(0)} \right)^2 \right] = k_4^2 \left( b_0^2 I_x \right)$, hence $k_4 = \left( b_0^2 I_x \right)^{-1/2}$.

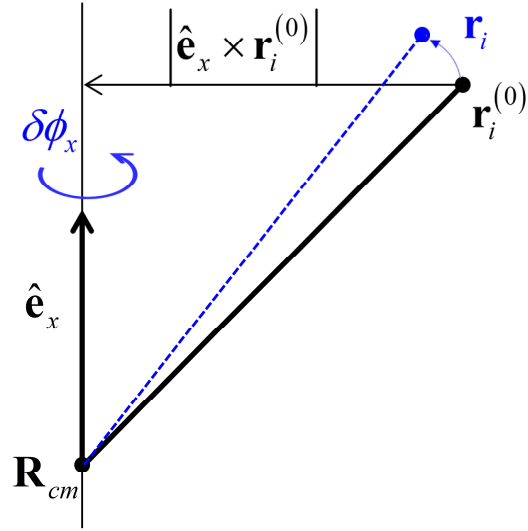

FIG. S4. Definition of displacement distance $\left|\mathbf{r}_i - \mathbf{r}_i^{(0)}\right|$ made by a small rotation $\delta\phi_x$ about the principal axis $\hat{\mathbf{e}}_x$.

The rotational modes $q_4$, $q_5$, $q_6$ correspond to rotations of $\vec{r}_0$ by small angles $\delta\phi_\nu$ about one of the principal axes $\hat{\mathbf{e}}_x$, $\hat{\mathbf{e}}_y$, $\hat{\mathbf{e}}_z$. For such a rotation, points $\mathbf{r}_i^{(0)}$ are displaced to $\mathbf{r}_i$. For example, for a rotation about $\hat{\mathbf{e}}_x$ by $\delta\phi_x$ one obtains from Eq. (21)

$$\delta q_4 = k_4 \sum_{i=1}^{N} \left(0, -z_i^{(0)}, y_i^{(0)}\right) \cdot \left(\mathbf{r}_i - \mathbf{r}_i^{(0)}\right) = k_4 \sum_{i=1}^{N} \left(\hat{\mathbf{e}}_x \times \mathbf{r}_i^{(0)}\right) \cdot \left(\mathbf{r}_i - \mathbf{r}_i^{(0)}\right)$$
$$= k_4 \sum_{i=1}^{N} \sqrt{\left(y_i^{(0)}\right)^2 + \left(z_i^{(0)}\right)^2} \left|\mathbf{r}_i - \mathbf{r}_i^{(0)}\right| = k_4 \sum_{i=1}^{N} \left[\left(y_i^{(0)}\right)^2 + \left(z_i^{(0)}\right)^2\right] \delta\phi_x . \quad (32)$$
$$= k_4 \left(b_0^2 I_x\right) \delta\phi_x = b_0 \sqrt{I_x} \delta\phi_x$$

In the second line we use that $\hat{\mathbf{e}}_x \times \mathbf{r}_i^{(0)}$ points in the direction of $\mathbf{r}_i - \mathbf{r}_i^{(0)}$ (for small $\delta\phi_x$) and that $\left|\hat{\mathbf{e}}_x \times \mathbf{r}_i^{(0)}\right|$ equals the distance $\sqrt{\left(y_i^{(0)}\right)^2 + \left(z_i^{(0)}\right)^2}$ of $\mathbf{r}_i^{(0)}$ from $\hat{\mathbf{e}}_x$. Furthermore, $\left|\mathbf{r}_i - \mathbf{r}_i^{(0)}\right| = \sqrt{\left(y_i^{(0)}\right)^2 + \left(z_i^{(0)}\right)^2} \delta\phi_x$ (Fig. S4). In the last line of Eq. (32) we use Eq. (31) and $I_x = b_0^2 \Gamma_{11}$ with $\Gamma_{\mu\nu}$ from Eq. (30). Similar calculations for rotations about $\hat{\mathbf{e}}_y$, $\hat{\mathbf{e}}_z$ yield

$$dq_4 = b_0 \sqrt{I_x} d\phi_x , \quad dq_5 = b_0 \sqrt{I_y} d\phi_y , \quad dq_6 = b_0 \sqrt{I_z} d\phi_z . \quad (33)$$

The infinitesimal rotations in (33) may be extended to finite rotations by parameterizing a finite rotation by Euler angles using the XYZ convention (corresponding to subsequent infinitesimal rotations about the principal axes $\hat{\mathbf{e}}_x$, $\hat{\mathbf{e}}_y$, $\hat{\mathbf{e}}_z$, see Eq. (6.1) in reference [15]). The rotational degrees of freedom of the molecule thus amount to a contribution

$$\int dq_4 \int dq_5 \int dq_6 = b_0^3 \sqrt{I_1 I_2 I_3}\, 8\pi^2 \tag{34}$$

where $8\pi^2 = 2\pi\,\Omega$ with the surface area $\Omega = 4\pi$ of the three-dimensional unit sphere. Combining Eqs. (26) and (34) we obtain $\prod_{n=1}^{6} \int dq_n = V N^{3/2} b_0^3 \sqrt{I_1 I_2 I_3}\, 8\pi^2$, and inserting in Eq. (23) yields Eq. (18).

The occurrence of the principal moments of inertia[a] $I_x$, $I_y$, $I_z$ in (18) can be understood as follows. The molecule's conformations described by Cartesian coordinates in Eq. (1) are discretized on a uniform lattice of lattice constant $a$. For a rotation of the molecule by a given angle $\phi$ about some axis, the arc length traced out by an atom in the molecule in terms of $a$ depends on the distance of the atom from the rotation axis. An atom that is farther away from the axis amounts to more conformations than an atom that is closer. The cumulative effect of this dependence gives rise to the moment-of-inertia factor for the given rotation axis.

## SVII. TI-NMA Calculations

In TI-NMA the absolute free energy $F$ for the target system (at $\lambda = 0$) is calculated as

$$F = F_0 - \Delta F_{TI} \tag{35}$$

where $F_0$ is the ground-state free energy obtained from NMA (Section SVI) and $\Delta F_{TI} = \Delta F_1 + \Delta F_2$ is the free energy difference obtained from both phases of TI (Section SV). Table S2 shows thermodynamic quantities obtained by TI-NMA for all knotted configurations studied in this work. The free energy of each knot type is compared with the equilibrium distributions (ESP, Section SIII) to verify the accuracy of the TI-NMA method (cf. main text, Fig. 5).

TABLE SII. Free Energies of knotted chains obtained by TI-NMA

| Knot-Type | $L/P = 20$, $N = 100$ | | | $L/P = 40$, $N = 200$ | | |
|---|---|---|---|---|---|---|
| | $\beta F_0(\lambda = 200)$ | $\beta F(\lambda = 0)$ | $\beta \langle U \rangle(\lambda = 0)$ [a] | $\beta F_0(\lambda = 200)$ | $\beta F(\lambda = 0)$ | $\beta \langle U \rangle(\lambda = 0)$ |
| $0_1$ | 1080.05 | 298.16(0.05)[b] | 151.14(0.10) | 2165.24 | 585.40(0.06) | 300.61(0.13) |
| $3_1$ | 1077.09 | 304.89(0.03) | 152.59(0.17) | 2162.22 | 590.53(0.11) | 301.14(0.13) |
| $4_1$ | 1077.85 | 307.77(0.08) | 154.08(0.19) | 2162.91 | 592.98(0.16) | 301.38(0.30) |
| $5_1$ | 1076.80 | 311.10(0.08) | 154.47(0.10) | 2162.02 | 595.31(0.13) | 301.59(0.32) |
| $5_2$ | 1078.28 | 311.19(0.15) | 154.84(0.13) | 2163.45 | 595.35(0.09) | 301.65(0.13) |
| $6_1$ | 1077.43 | 314.84(0.21) | 156.20(0.32) | 2162.90 | 598.14(0.18) | 302.02(0.20) |
| $6_2$ | 1078.12 | 314.25(0.15) | 156.14(0.28) | 2163.12 | 598.34(0.17) | 301.52(0.28) |
| $6_3$ | 1079.48 | 315.44(0.38) | 156.57(0.36) | 2163.78 | 598.24(0.24) | 302.40(0.18) |

[a] Mean energy of the unconstrained system obtained from simulation at $\lambda = 0$
[b] Values in parentheses represent SEM (see text)

Values of $F_0(\ell, L; P)$ and $\Delta F_{TI}(\ell, L; P)$ were calculated for dual-loop sc models (Fig. S5). For fixed overall length $L$, we fit $F_0(\ell, L; P)$ and $\Delta F_{TI}(\ell, L; P)$ to smooth functions of $\ell$ to obtain the free energy $F(\ell, L; P) = F_0 - \Delta F_{TI}$. A smooth function is fit to the data points by cubic spline interpolation (red line). The blue circles show the free energy change $\Delta F_{TI}(\ell, L; P)$ obtained by TI for $\lambda = 0 \rightarrow 200$. For every value of $\lambda$, $\eta = 5 \times 10^5$ conformations were generated for chains with $N < 60$ segments and $\eta = 10^5$ conformations were generated for chains with $N \geq 60$ segments. The error was estimated by dividing each large ensemble into 5 smaller ensembles and calculating the SEM from the smaller ensembles. The obtained values for $\Delta F_{TI}(\ell, L; P)$ were interpolated with a polynomial of the form $f(x) = c_1 + c_2 x^2 + c_3 x^4 + c_4 x^6$ where $x = \ell/L - 0.5$ (blue line). To obtain results independent of the length $a$ (Section SI) we consider the free energy difference $\Delta F = F(\ell, L; P) - F(L/2, L; P)$ (Fig. S6).

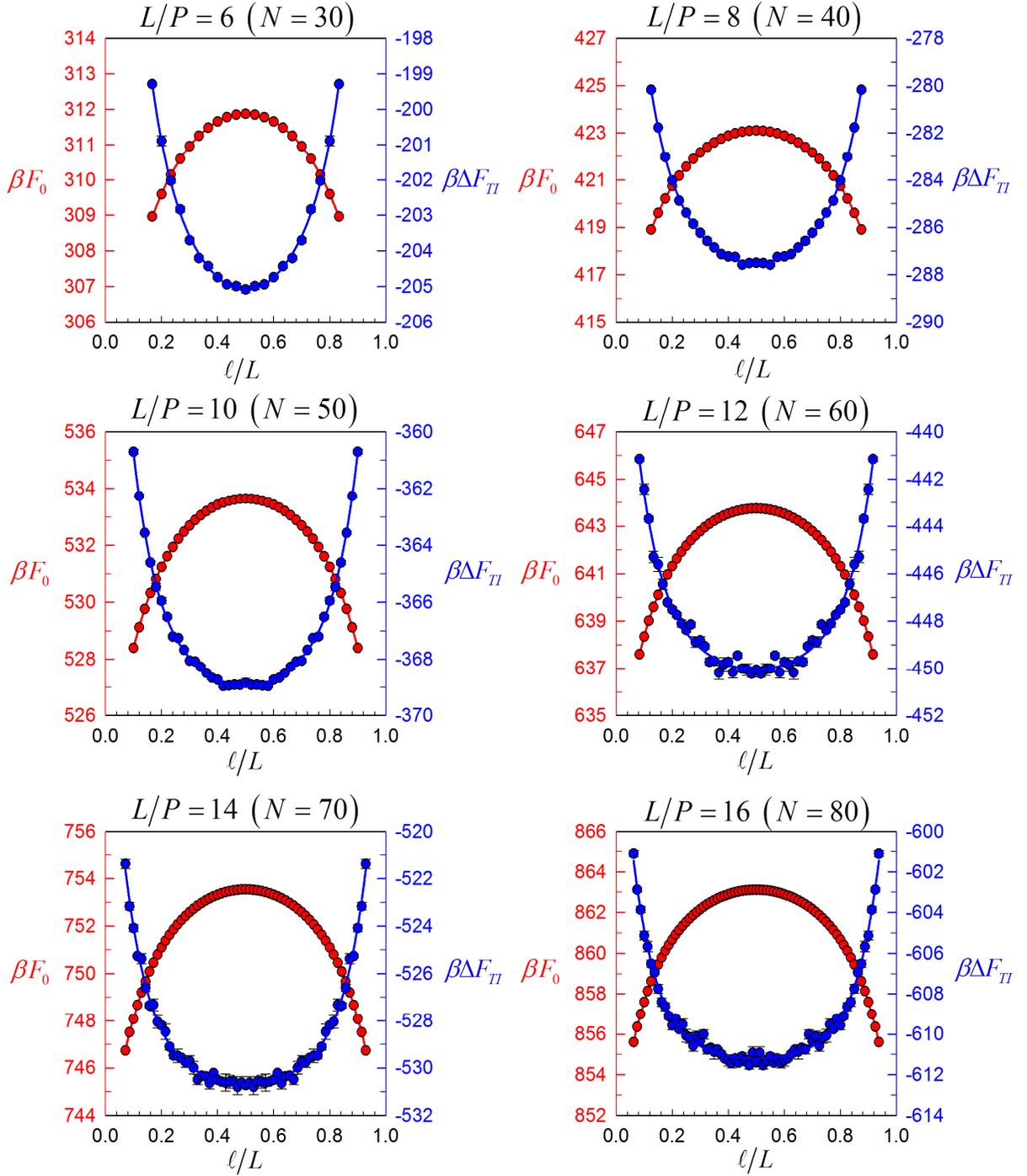

FIG. S5. Free energies for dual-loop sc models obtained by TI-NMA. Each data point represents the free energy of a molecule with fixed length ratio $L/P$ and loop length $\ell$. Red circles show $F_0(\ell, L; P)$ obtained by NMA at $\lambda = 200$. (See text.)

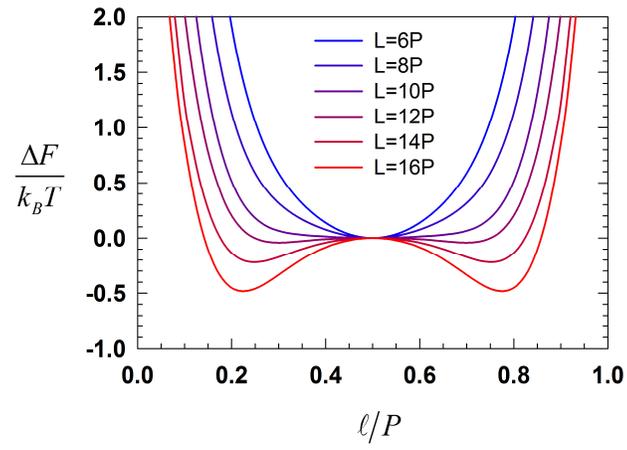

FIG. S6. Free energies $\Delta F = F(\ell, L; P) - F(L/2, L; P)$ of dual-loop sc models as a function of loop length $\ell$ for given total contour length $L$.

**REFERENCES**

[1] A. V. Vologodskii, S. D. Levene, K. V. Klenin, M. Frank-Kamenetskii and N. R. Cozzarelli, J. Mol. Biol. **227** (4), 1224-1243 (1992).

[2] H.-P. Hsu, W. Paul and K. Binder, Macromolecules **43** (6), 3094-3102 (2010).

[3] D. Shore, J. Langowski and R. L. Baldwin, Proc. Natl. Acad. Sci. U.S.A. **78** (8), 4833-4837 (1981).

[4] S. B. Smith, L. Finzi and C. Bustamante, Science **258** (5085), 1122-1126 (1992).

[5] C. Bouchiat, M. D. Wang, J. F. Allemand, T. Strick, S. M. Block and V. Croquette, Biophys. J. **76** (1), 409-413 (1999).

[6] V. V. Rybenkov, N. R. Cozzarelli and A. V. Vologodskii, Proc. Natl. Acad. Sci. U.S.A. **90** (11), 5307-5311. (1993).

[7] A. A. Podtelezhnikov, N. R. Cozzarelli and A. V. Vologodskii, Proc. Natl. Acad. Sci. U.S.A. **96** (23), 12974-12979 (1999).

[8] G. Gouesbet, S. Meunier-Guttin-Cluzel and C. Letellier, Appl. Math. Comput. **105** (2–3), 271-289 (1999).

[9] P. Ramadevi, T. R. Govindarajan and R. K. Kaul, Mod. Phys. Lett. A **09** (34), 3205-3217 (1994).

[10] R. G. Scharein, The University of British Columbia, 1998.

[11] J. Nocedal and S. J. Wright, *Numerical Optimization*, 2nd ed. (Springer, New York, 2009).

[12] S. A. Harris, E. Gavathiotis, M. S. Searle, M. Orozco and C. A. Laughton, J. Am. Chem. Soc. **123** (50), 12658-12663 (2001).

[13] Q. Cui and I. Bahar (Eds.), *Normal mode analysis: theory and applications to biological and chemical systems*. (Chapman & Hall/CRC, Boca Raton, 2006).

[14] H. Goldstein, C. P. Poole and J. L. Safko, *Classical Mechanics*, 3rd ed. (Addison Wesley, San Francisco, 2001).

[15] A. Hanke, S. M. Giovan and S. D. Levene, Prog. Theor. Phys. Supplement **191**, 109-129 (2011).